\documentclass[a4paper,fleqn,usenatbib]{mnras}

\usepackage{newtxtext,newtxmath}
\usepackage{xcolor}
\usepackage[T1]{fontenc}
\usepackage{ae,aecompl}

\usepackage{graphicx}	
\usepackage{amsmath}	
\usepackage{siunitx}
\usepackage{subfig}
\usepackage[normalem]{ulem} 
\usepackage{soul}

\newcommand\Tstrut{\rule{0pt}{1.1ex}}       

\newcommand{\ngc}{NGC\,5548}	            
\newcommand{\se}{[S\,\textsc{viii}]}	    
\newcommand{\sn}{[S\,\textsc{ix}]}	        
\newcommand{\sit}{[Si\,\textsc{x}]}	        
\newcommand{\silvi}{{[Si\,\textsc{vi}]}}    
\newcommand{\paa}{Pa\,$\upalpha$}	    
\newcommand{\pab}{Pa\,$\upbeta$}	    
\newcommand{\pad}{Pa\,$\updelta$}	    
\newcommand{\pae}{Pa\,$\upepsilon$}	    
\newcommand{\hi}{H\,\textsc{i}}	        
\newcommand{\heii}{He\,\textsc{ii}}	    
\newcommand{\ci}{[C\,\textsc{i}]}	    
\newcommand{\sii}{[S\,\textsc{ii}]}	    
\newcommand{\siii}{[S\,\textsc{iii}]}	
\newcommand{\oiii}{[O\,\textsc{iii}]}	
\newcommand{\feii}{[Fe\,\textsc{ii}]}	
\newcommand{\fevii}{[Fe\,\textsc{vii}]}	
\newcommand{\fex}{[Fe\,\textsc{x}]}	
\newcommand{\brd}{Br\,$\updelta$}	    
\newcommand{\brg}{Br\,$\upgamma$}	    
\newcommand{\htwo}{H$_2$}	                
\newcommand{\um}{\SI{}{\micro\meter}}            
\newcommand{\kms}{{km\,s$^{-1}$}}           
\newcommand{\cloudy}{\textsc{cloudy}}           
\newcommand{\comment}[1]{}

\title[Near-IR coronal lines in NGC\,5548]{Multiple locations of near-infrared coronal lines in NGC\,5548}

\author[D. Kynoch et al.]{Daniel Kynoch,$^{1,2}$\thanks{E-mail: daniel.kynoch@asu.cas.cz}\thanks{Visiting Astronomer at the Infrared Telescope Facility, which is operated by the University of Hawaii under contract NNH14CK55B with the National Aeronautics and Space Administration (NASA)}
Hermine Landt,$^2$\footnotemark[2]\thanks{E-mail: hermine.landt@durham.ac.uk}
Maryam Dehghanian,$^3$
Martin J. Ward$^2$\footnotemark[2]
\newauthor
and Gary J. Ferland$^3$
\\
$^{1}$Astronomical Institute, Czech Academy of Sciences, Bo\v{c}n\'{i} II 1401, 141 00 Prague, Czech Republic \\
$^2$Centre for Extragalactic Astronomy, Department of Physics, Durham University, South Road, Durham, DH1 3LE, UK \\
$^3$Department of Physics and Astronomy, University of Kentucky, Lexington, KY 40506, USA \\
}

\date{Accepted XXX. Received YYY; in original form ZZZ}

\pubyear{2022}

\begin{document}
\label{firstpage}
\pagerange{\pageref{firstpage}--\pageref{lastpage}}
\maketitle

\begin{abstract}
We present the first intensive study of the variability of the near-infrared coronal lines in an active galactic nucleus (AGN). We use data from a one-year long spectroscopic monitoring campaign with roughly weekly cadence on NGC\,5548 to study the variability in both emission line fluxes and profile shapes. We find that in common with many AGN coronal lines, those studied here are both broader than the low-ionisaton forbidden lines and blueshifted relative to them, with a stratification that implies an origin in an outflow interior to the standard narrow line region. We observe for the first time \se\ and \silvi\ coronal line profiles that exhibit broad wings in addition to narrow cores, features not seen in either \sn\ or \sit. These wings are highly variable, whereas the cores show negligible changes. The differences in \textcolor{black}{both the} profile shapes and variability \textcolor{black}{properties} of the different line components indicate that there are at least two coronal line regions in AGN. We associate the variable, broad wings with the base of an X-ray heated wind evaporated from the inner edge of the dusty torus. The coronal line cores may be formed at several locations interior to the narrow line region\textcolor{black}{: either} along this accelerating, clumpy wind or in the much more compact outflow identified with the obscurer and so emerging on scales similar to the outer accretion disc and broad line region.
\end{abstract}




\section{Introduction}
\label{sec:intro}
Active galactic nuclei (AGN) display emission lines from both permitted and forbidden transitions. The latter are usually associated with the narrow emission line region (NLR), which is formed by gas of low densities ($\log[n_{\rm H}/\mathrm{cm}^{-3}] \sim 3$--6) located at relatively large distances from the central ionising source (from a few pc up to several 100~pc). But some of these forbidden emission lines require energies $\gtrsim 100$~eV to form the corresponding ions, have higher critical densities for collisional deexcitation ($\log[n_{\rm e}^{\rm crit}/\mathrm{cm}^{-3}] \sim 7$--10) and broader profiles (full widths at half maxima [FWHM] $\sim 500-1500$~km~s$^{-1}$) than the low-ionisation narrow emission lines \citep[e.g.,][]{Pen84, App88, Gian95, Erk97, Rod02, Rod11}: these are the so-called `coronal lines', named after their presence in the spectrum of the solar corona \citep{Oke68}.
Coronal lines are forbidden fine-structure transitions arising from highly-ionised states of heavy metals. Low-energy electrons or weak interactions with high-energy electrons can efficiently excite these transitions from the ground state, however, the formation of the ions themselves requires relatively high energies.

Both types of AGN display coronal lines \citep{Ost77, Koski78}, but they are stronger in broad-line (type~1) than in narrow-line (type~2) AGN relative to their low-ionisation narrow emission lines \citep{Mur98}. Therefore, it is likely that the coronal line region has two components, one compact and one spatially extended, with only the latter remaining unobscured by the dusty torus in type~2 AGN. Support for this assumption comes from the fact that the emission from this region is often extended but much less so than that from the low-ionisation NLR (on scales of $\sim 80-150$~pc; e.g.\ \citealt{Prieto05, Mueller06, Mueller11, Maz13, Riffel21}; although \citealt{Negus21} have recently reported coronal line emission on kpc scales).
Furthermore, coronal lines are often blueshifted relative to the low-ionisation narrow lines \citep[e.g.,][]{Pen84, Erk97, Rod02}, which indicates an outflowing wind component. Given their similar physical conditions, it could be that the partly ionised gas that produces absorption lines and edges in the soft X-ray spectra of AGN, i.e., the so-called `warm absorber', produces also the coronal lines in its (colder) outer regions \citep{Netzer93, Erk97, Por99}. In any case, the coronal line region is most likely dust free, since strong emission from refractory elements such as iron, silicon and calcium are observed, which would be severely reduced in a dusty environment \citep{Ferg97}.

The high ionisation potentials ($\chi$) required for the coronal lines can be found either in a hot, collisionally ionised plasma or be produced by the hard continuum in AGN if the gas is photoionised. In the first case, the electron temperatures would be of the order of $\log(T_{\rm e}/\mathrm{K}) \approx6$ and in the second case much lower ($\log[T_{\rm e}/\mathrm{K}] \sim 4$--5). Currently, photoionisation is favoured, since for most AGN the observed flux ratios between different coronal lines can be reproduced within a factor of $\sim 2-3$ by these models \citep{Oliva94, Ferg97, Landt15a, Landt15b}, whereas in the case of a hot plasma either its temperature needs to be fine-tuned within a very narrow range \citep{Oliva94} or no acceptable fit can be obtained \citep{Landt15a, Landt15b}.

More recently, growing interest in near-IR coronal lines has arisen due to their potential for yielding estimates of the black hole mass in AGN \citep{Cann18, Rod20}. Since these coronal lines can potentially probe the accretion disc spectral energy distribution (SED) at far-UV/soft-X-ray energies, a regime difficult to observe but most influenced by the mass of the black hole, they could uncover the long-sought population of intermediate black holes, which are crucial for our understanding of black hole growth over cosmic time \citep{Hopkins12}. Even more importantly, if near-IR coronal lines can help identify AGN in dwarf galaxies in large numbers \citep{Bohn21, Cann21}, it would open up a unique opportunity to understand the role of AGN feedback in galaxy evolution and to test cosmological models. Since dwarf galaxies are believed to be dark matter-dominated, these sources serve as important probes in the low-mass halo regime in particular for the Lambda Cold Dark Matter ($\Lambda$CDM) paradigm \citep{Nav19}.

Variability studies can strongly constrain the properties of the coronal line region, in particular if several lines can be studied simultaneously and together with other AGN components such as the broad emission line region (BLR) and the UV/X-ray continuum. However, since these emission lines \textcolor{black}{are relatively weak} and so require high-quality spectroscopy, very few studies of this kind have been attempted so far. \citet{Vei88} presented the only systematic study of coronal line variability. In a sample of $\sim 20$ AGN, he found firm evidence that both the \fevii~$\lambda 6087$ and \fex~$\lambda 6375$ emission lines varied (during a period of a few years) for only one source (NGC\,5548) and tentative evidence for another seven sources (including NGC~4151). Then, within a general optical variability campaign on Mrk~110 for about half a year, \citet{Kolla01} reported strong \fex~variations. Strong variability of the coronal lines manifested mainly as a fading of the flux was first reported for IC~3599 \citep{Grupe95, Brandt95} and is now usually associated with a new class of non-active galaxies, the so-called `strong coronal line emitters' (or `coronal line forest AGN'). Most of these sources have been detected in the Sloan Digital Sky Survey (SDSS: \citealt{York00}) and a stellar tidal disruption event seems to be the most plausible explanation for the strong fading of their coronal lines over a time period of several years \citep{Kom08, Gel09, Kom09, WangT11, WangT12, Yang13, Rose15, Winkler16, Cer21, VV21}.

\citet{Landt15a} and \citet{Landt15b} presented the first extensive studies of the coronal line variability in individual AGN. Their data sets for the nearby, well-known sources NGC~4151 and NGC\,5548 included a handful of epochs of quasi-simultaneous optical, near-IR and X-ray spectroscopy spanning a period of several years. They found very different variability behaviours for the two sources, with only weak variations detected for the coronal lines in NGC~4151, but strong flux variability (mainly a decrease) by factors of $\sim 2-4$ observed in NGC\,5548. In both sources, the coronal line gas density was constrained to relatively low values of $\log(n_{\rm e}/\mathrm{cm}^{-3}) \sim3$ for a relatively high ionisation parameter of $\log U \sim 1$, which put it at a distance from the central ionising source of a few light~years and so well beyond the hot inner face of the obscuring dusty torus. Therefore, they proposed that the coronal line region in AGN is an independent entity rather than part of a continuous gas distribution connecting the BLR and low-ionisation NLR, possibly an X-ray heated wind as first suggested by \citet{Pier95}.

Here we revisit the variability of the near-IR coronal lines in \ngc\ with a much improved data set that allows us to study in detail both flux and profile shape variations. Our paper is structured as follows. In Section \ref{data}, we briefly discuss the near-IR spectroscopy, which we analyse in detail in Section~\ref{analysis}. In Section~\ref{results}, we present the results on the coronal line profiles and their variability, which we compare to theoretical photoionisation simulations in Section~\ref{sec:cloudy}. In Section~\ref{sec:discussion}, we discuss the likely origin of the near-IR coronal lines in \ngc.  Finally, in Section \ref{sec:conclusions}, we present a summary of our main results and conclusions. We quote all laboratory line wavelengths as vacuum wavelengths and define velocities as negative if they are in the blue-shifted (outflowing) direction.

\section{The data} \label{data}

\citet{Landt19} observed NGC~5548 between 2016 August and 2017 July with the SpeX spectrograph \citep{Rayner03} at the NASA Infrared Telescope Facility (IRTF), a 3~m telescope on Maunakea, Hawaii. The main aim of this near-IR spectroscopic reverberation mapping campaign was to measure the time delay of the hot dust in the obscuring torus together with estimates of the torus radius based on thermal equilibrium arguments. Another important goal was to study the variability of emission lines such as those from the coronal line region. 

The campaign achieved a total of 18 near-IR spectra with an average cadence of about ten days, excluding a 3.5-month period when the source was unobservable. They used the short cross-dispersed (SXD) mode (0.7--\SI{2.55}{\micro\meter}) and a $0.3^{\prime\prime} \times 15^{\prime\prime}$ slit oriented at the parallactic angle, resulting in an average spectral resolution of $R=2000$ or full width at half maximum (FWHM) $\sim 150$~km~s$^{-1}$. The spectra have in general a relatively high signal-to-noise (S/N) ratio with an average continuum $\mathrm{S/N} \sim 100$.

Care was taken to ensure \textit{a posteriori} an accurate absolute flux calibration. As described in detail in \cite{Landt19} (see their section~3.3), they used the narrow forbidden emission line \siii~$\lambda 9531$ to align the flux scale of the spectra and verified it with a photometric light-curve. Furthermore, the impact of the extended low-ionisation emission line region on the enclosed flux in the slit was assessed based on a near-IR Integral Field Unit (IFU) observation. For our following analysis, we used the observed spectra with their multiplicative photometric correction factors applied. We estimated that the uncertainty on the wavelength calibrations of our spectra is on average 0.15~\AA, which corresponds to $\approx2-5$~\kms\ at the location of the lines studied. We report velocity shifts of the coronal lines measured relative to the \siii~$\lambda 9531$ narrow emission line and the wavelength calibration uncertainty was added to the measurement uncertainties.

\section{The spectral analysis} \label{analysis}

\begin{table}
\centering
\caption{Coronal emission lines and their contaminants}
\label{tab:contaminants}
\begin{tabular}{lccclc}
\hline
\multicolumn{4}{c}{Coronal emission line} & \multicolumn{2}{c}{Contaminant emission line} \\
Ion & $\lambda$ & $\chi$ & $\log\left(n_\mathrm{e}^\mathrm{crit}\right)$ & Ion & $\lambda$ \\
&  [\um] & [eV] & [cm$^{-3}$] & & [\um] \\
(1) & (2) & (3) & (4) & (5) & (6)  \\
\hline
\Tstrut
\se      & 0.9914 & 281.0 &  9.5 &  \ci\              & 0.9827 \\
         &        &       &      &  \ci\              & 0.9853 \\      
         &        &       &      &  \hi\ (\pad)       & 1.005   \\
         &        &       &      &  \heii\            & 1.0126 \\
\sn      & 1.2523 & 328.8 &  9.4 &  Unknown           & 1.248   \\
         &        &       &      &  \feii\            & 1.2570 \\
         &        &       &      &  Unknown           & 1.262   \\
         &        &       &      &  \hi\ (\pab)       & 1.282   \\
\sit     & 1.4305 & 351.1 &  8.1 &                    &        \\
\silvi   & 1.9650 & 166.8 &  8.5 &  \hi\ (\paa)       & 1.875   \\
         &        &       &      &  \hi\ (\brd)       & 1.9440  \\
         &        &       &      &  \htwo\            & 1.957   \\
\hline
\end{tabular}

\parbox[]{\columnwidth}{The columns are: (1) Ion; (2) vacuum rest-frame wavelength; (3) ionisation potential; and (4) critical density calculated with \cloudy\ for the coronal emission lines at temperature $\log(T_\mathrm{e}/\mathrm{K})=4$; (5) ion; and (6) vacuum rest-frame wavelength for the contaminating broad and narrow emission lines. The emission line parameters of all these lines are listed in Table \ref{tab:mean_fits}.}

\end{table}

The large wavelength range our cross-dispersed near-IR spectra covers four strong coronal lines in \ngc. Two of them are produced by highly-ionised sulphur (\se~and \sn) and two by highly-ionised silicon (\silvi~and \sit). We list their basic properties in Table~\ref{tab:contaminants}. Our main aim was to reliably isolate the coronal lines in the individual spectra in order to measure their fluxes, velocity shifts and profile shapes and study their variability. However, since coronal lines are in general weak emission lines, we analysed them first in the high-quality mean spectrum presented by \citet{Landt19} and then used these results as a guide for our analysis of the individual spectra, which have a lower $S/N$ in the wavelength regions of interest. 

We performed the emission line and continuum fits using custom Python scripts employing the package \textsc{lmfit} \citep{Newville14}, which is a non-linear, least-squares, curve-fitting routine based on the Levenberg-Marquardt algorithm.  \textcolor{black}{The majority of the emission lines can be adequately modelled with Gaussian profiles. 
From these fits we have calculated the emission line fluxes and FWHMs and the associated uncertainties.
All narrow emission lines were allowed to have widths varying in the range of $\mathrm{FWHM}=100$--900~\kms\
and the widths of broad Gaussian components were allowed to vary in the range of 1000--9999~\kms.
Values quoted for the FWHMs of the emission lines have been corrected for instrumental broadening.
The velocity shifts of the near-IR lines are determined from the centroids of the fitted Gaussians and measured relative to the \siii$\lambda9531$ line, which is assumed to have zero velocity shift.
The uncertainties on the velocity shifts incorporate the uncertainty on the wavelength calibration of our spectra.
More details on the fitting procedures for specific lines and blends are given in the following subsections.} 

\subsection{The mean spectrum} \label{sec:meanspec}

\begin{figure*}
	\includegraphics[width=1.5\columnwidth]{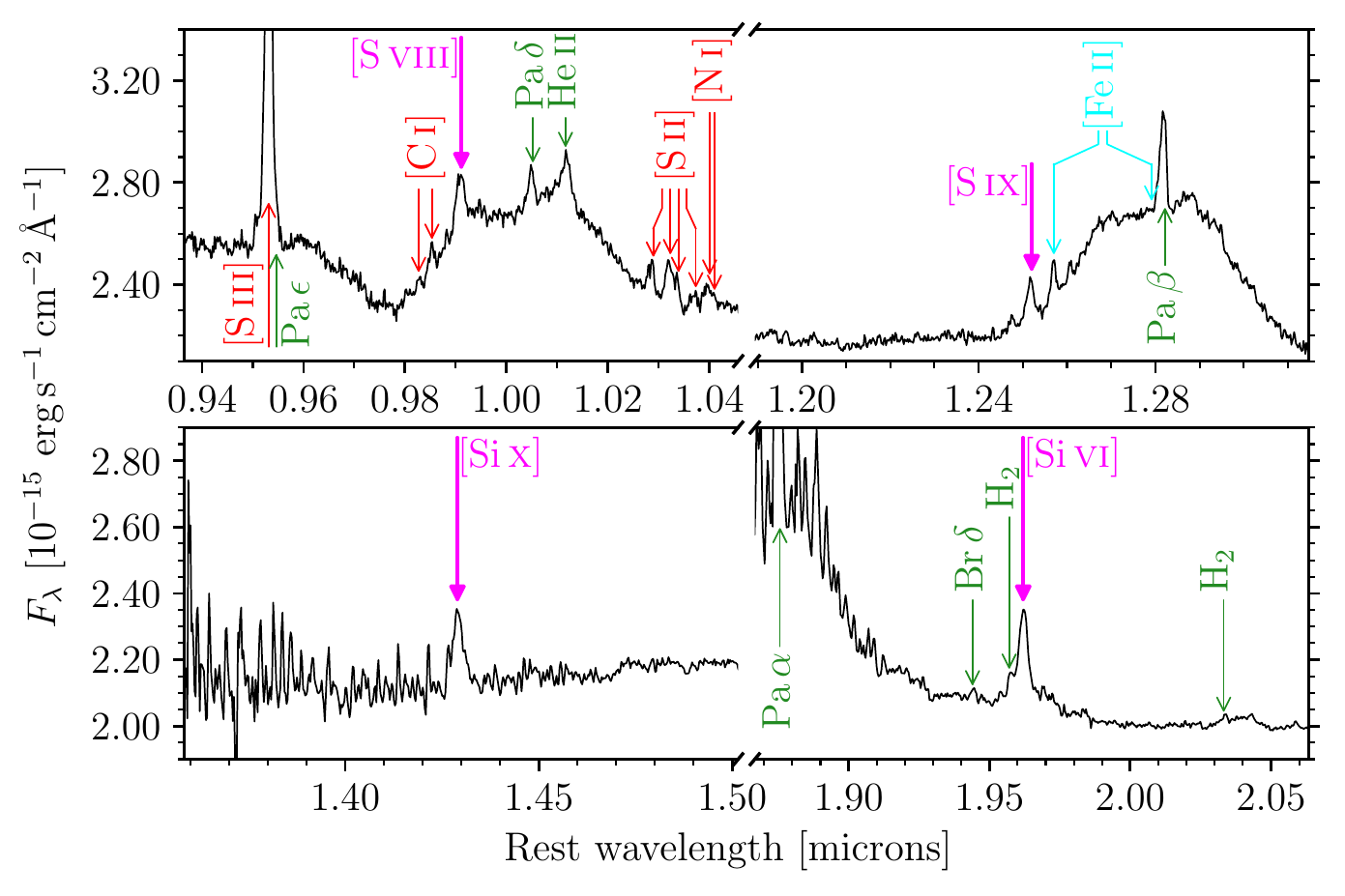}
    \caption{Spectral regions in the mean spectrum around the four near-IR coronal lines studied. The coronal lines are labelled in magenta, whereas contaminating permitted and forbidden transitions are labelled in green and red, respectively. Forbidden transitions of iron are labelled in cyan. The wavelengths and fluxes of all the marked emission lines are listed in Table \ref{tab:mean_fits}.}
    \label{fig:fourlines}
\end{figure*}
\begin{table*}
	\centering
	\caption{Emission line parameters measured in the mean spectrum}
	\label{tab:mean_fits}
	\begin{tabular}{cccccc} 
\hline
Ion & $\lambda$ & Line & Velocity & FWHM & Flux \\
& [\um] & component & offset & [km\,s$^{-1}$] & [$10^{-15}$~erg\,s$^{-1}$\,cm$^{-2}$] \\
(1) & (2) & (3) & (4) & (5) & (6) \\
\hline
\siii       & 0.90711 &           &                   & $378\pm11$     & $13.3\pm0.5$ \\
\siii       & 0.95332 &           &                   & $365\pm7$      & $33\pm3$ \\
\ci         & 0.98268 &           & $+48\pm24$        & $483\pm27$     & $0.5\pm0.2$ \\
\ci         & 0.98530 &           & $+48\pm24$        & $483\pm27$     & $2.1\pm0.2$ \\
\se\        & 0.99138 & core      & $-119\pm15^\star$ & $737\pm39$     & $7.3\pm0.4$ \\
"           &         & blue wing &                   &                & $1.7\pm0.4$ \\
"           &         & red wing  &                   &                & $2.8\pm0.7$ \\
\hi\ (\pad) & 1.00521 & narrow    & $-69\pm25$        & $461\pm26$     & $2.9\pm0.2$ \\
\hi\ (\pad) &         & broad~1   & $-2961$           & $5091$         &  \\
\hi\ (\pad) &         & broad~2   & $+1680$           & $5149$         & $109\pm11$ \\
\heii\      & 1.01264 & narrow    & $-7\pm18$         & $461\pm26$     & $2.9\pm0.2$ \\
\heii\      &         & broad     & $+1576\pm1025$    & $8847\pm956$   & $49\pm11$ \\
Unknown     & 1.2475  &           &                   & $334\pm13$     & $1.3\pm0.2$ \\ 
\sn\        & 1.2523  &           & $-120\pm14^\star$ & $554\pm24$     & $4.3\pm0.2$ \\
\feii       & 1.25702 &           & $+25\pm18$        & $334\pm13$     & $2.2\pm0.2$ \\ 
Unknown     & 1.2612  &           &                   & $380\pm15$     & $<0.45$ \\
\hi\ (\pab) & 1.28216 & narrow    & $-55\pm12^\star$  & $334\pm13$     & $6.4\pm0.3$ \\
\hi\ (\pab) &         & broad~1   & $-2961\pm125$     & $5091\pm197$   & \\
\hi\ (\pab) &         & broad~2   & $+1680\pm112$     & $5149\pm141$   & $215\pm7$ \\
\sit\       & 1.430   &           & $-138\pm14^\star$ & $734\pm43$     & $7.9\pm0.6$ \\
\hi\ (\brd) & 1.94509 & narrow    & $-69\pm41$        & $334$          & $0.9\pm0.2$ \\  
\hi\ (\brd) &         & broad~1   & $-2961$           & $5091$         &  \\ 
\hi\ (\brd) &         & broad~2   & $+1680$           & $5149$         & $39$ \\  
\silvi\     & 1.9650  & core      & $-384\pm14^\star$ & $607\pm25$     & $13.0\pm0.6$ \\
"           &         & blue wing &                   &                & $3.6\pm0.4$ \\
"           &         & red wing  &                   &                & $8.7\pm0.6$ \\
\htwo\      & 2.0332  &           & $-66\pm17^\star$  & $174\pm25$     & $0.8\pm0.1$ \\
\hi\ (\brg) & 2.16612 & narrow    & $0\pm22$          & $389\pm46$     & $1.3\pm0.2$ \\
\hi\ (\brg) &         & broad~1   & $-2961$           & $5091$         & \\
\hi\ (\brg) &         & broad~2   & $+1680$           & $5149$         & $43.3\pm0.3$ \\
\hline
\end{tabular}
	
\parbox[]{11cm}{The columns are: (1) Ion; (2) vacuum wavelength; (3) fitted emission line component; (4) velocity offset of the emission line peak relative to \siii$\lambda0.95332$~\um: narrow line shifts with a significance $>3\sigma$ are marked $^\star$; (5) full width at half maximum of the emission line component corrected for an instrumental broadening of 150~\kms; and (6) integratedline flux. We give $1\sigma$ errors. We note that all \hi\ emission lines are assumed to have the same double-peaked broad-line profile as \pab\ and that we list for them only the total broad line flux.}
\end{table*}

\begin{figure*}
    \centering
    \includegraphics[width=1.9\columnwidth]{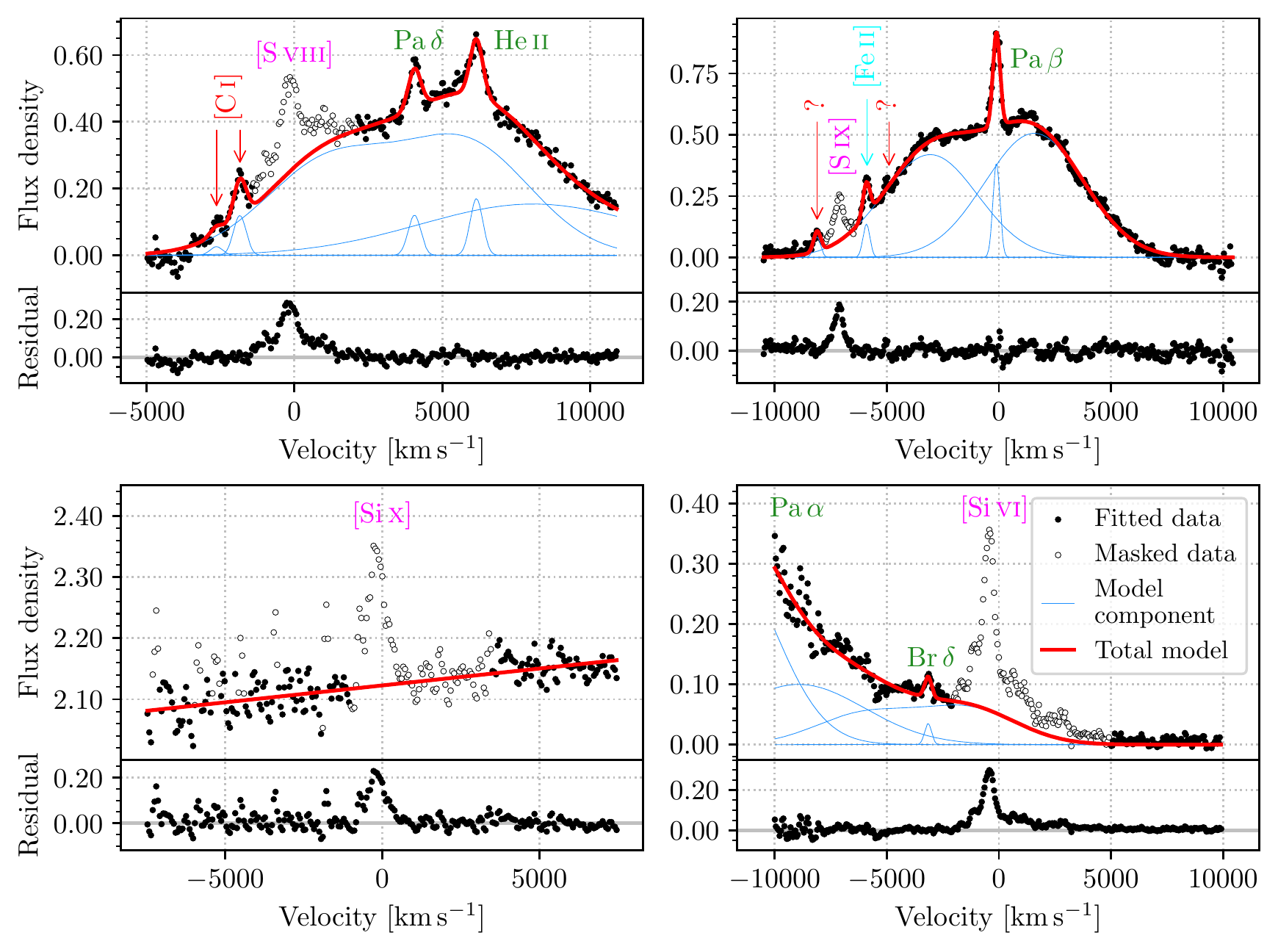}
    \caption[]{The decomposition of emission line blends around the four near-IR coronal lines using the mean spectrum. All coronal lines were masked out and not used in the fitting process. Individual emission lines are shown as blue dotted lines and the total models are shown as the red solid lines. The bottom panels for each coronal line show the flux residuals across the full emission line complex. All velocities are relative to the expected rest-frame wavelength of the coronal line, except in the case of \sn~where velocities are relative to \pab. See \textcolor{black}{Section~\ref{sec:meanspec} in the} text for more details.}
    \label{fig:mean_spec_fits}
\end{figure*}

Fig.~\ref{fig:fourlines} shows the wavelength regions around the four near-IR coronal lines in the mean spectrum. Out of these, only \sit\ is free from contaminating, neighboring emission lines, whereas the other coronal lines are located close to a hydrogen Paschen broad emission line and other chemical species (Table \ref{tab:contaminants}). Therefore, rather than isolating the coronal lines by `clipping' the profiles to the local continuum on the red and blue sides of the line, we carefully modelled all the line complexes around the coronal lines in order to deblend them from neighbouring features (Fig. \ref{fig:mean_spec_fits}). The resultant coronal line fluxes, their blends and other observed emission lines are listed in Table \ref{tab:mean_fits}. In the following we give details of the analysis.

\subsubsection{The continuum} \label{sec:continuum}
Firstly, the underlying pseudo-continuum is modelled and subtracted in four spectral regions: 0.7--1.22, 1.12--1.33, 1.39--1.53 and 1.65--2.4~\um. The 1.12--1.33 and 1.39--1.53~\um\ regions could be satisfactorily fit with a simple power-law, whereas the 0.7--1.22 and 1.65--2.4~\um\ regions required polynomial models to reproduce the curvature of the pseudo-continuum.

There are no other emission lines in the vicinity of \sit, therefore its profile could be obtained by simply subtracting the local pseudo-continuum flux. Whilst \sit\ is free from other emission lines, this spectral region is affected by telluric absorption, particularly on the blue side of the line. 
We were therefore careful to avoid the inclusion of residual telluric features when modelling the pseudo-continuum (see Fig.~\ref{fig:mean_spec_fits}). 
The coronal lines \se, \sn\ and \silvi\ are all blended with other emission lines. We determined their profiles by modelling the emission line complexes as further described below.

\begin{figure}
    \centering
    \includegraphics[width=\columnwidth]{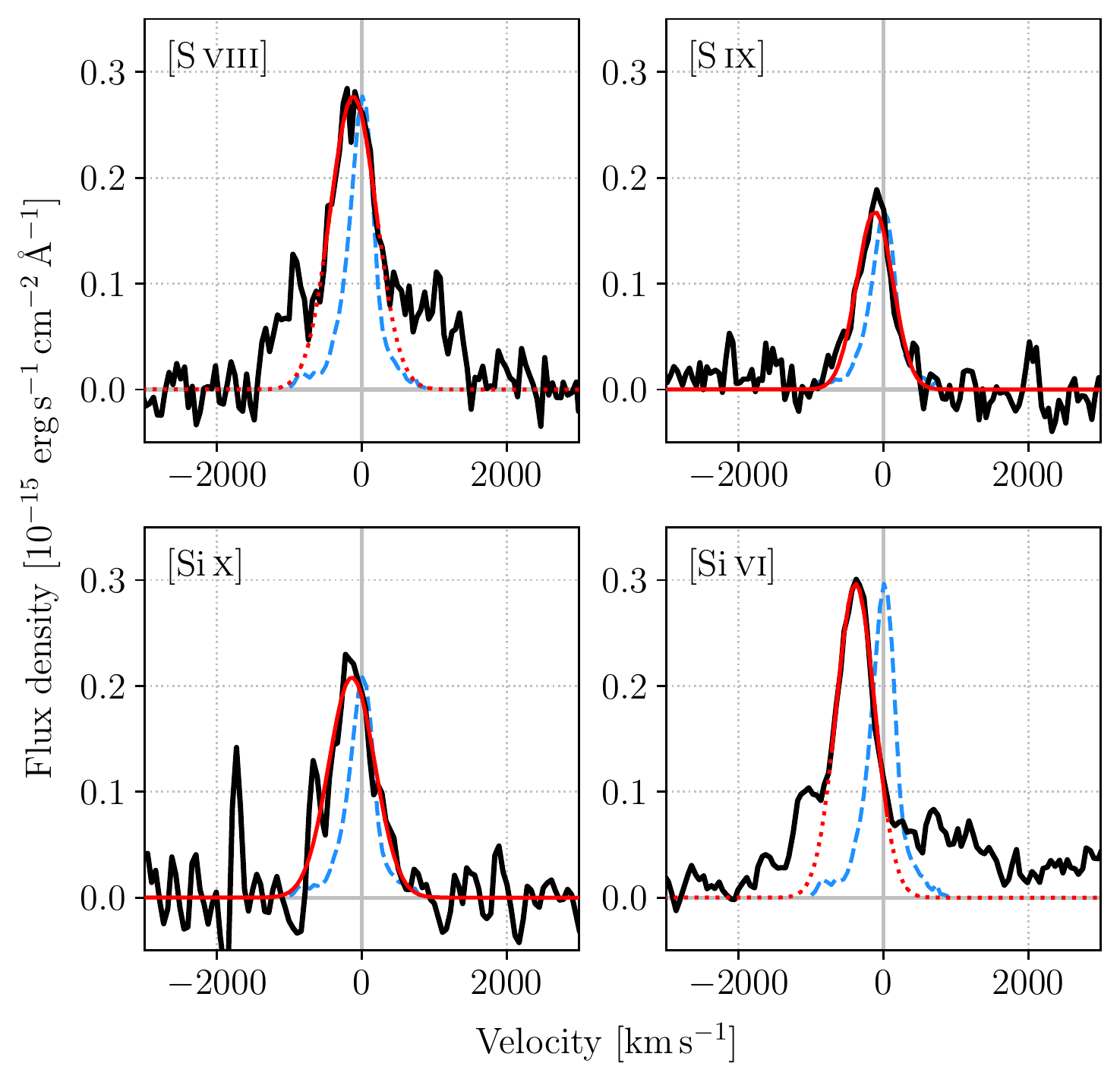}
    \caption{The coronal line profiles as extracted from the mean spectrum (black) and fit with Gaussians (red). 
    The total profiles of the \sn\ and \sit\ lines are fit with a single Gaussian, whereas only the central cores of the \se\ and \silvi\ lines are fit \textcolor{black}{(solid red lines). 
    For the latter two lines we additionally show the best fit extrapolated across the whole line profile (dotted red lines);} 
    blue and red excess flux is evident on these lines.
    For comparison, the \siii$\lambda9531$ profile (normalised in amplitude to each coronal line profile) is plotted with a blue dashed line.
    The four coronal lines are broader than \siii\ and blueshifted with respect to it.}  
    \label{fig:coronal_line_Gauss}
\end{figure}

\subsubsection{The \sn~line} \label{sec:mean_s_ix}
The \sn\ coronal line is located on the blue wing of broad \pab. \cite{Schonell17} analysed a high-spectral resolution near-IR spectrum of \ngc\ obtained in 2012 with the Near-Infrared Integral Field Spectrometer (NIFS) at Gemini North. They reported that \pab\ has two kinematically distinct components producing a double-peaked broad emission line.
Therefore, we decomposed the \pab~emission line using two Gaussians, one red and one blue of the line centre. 

The widths of all Gaussians modelling the narrow lines were tied together. In addition to the narrow, forbidden \feii\ lines at 1.2570 and 1.2791~\um, we find two unidentified lines at $\approx1.2483$ and $\approx1.2612$~\um. The former unidentified line appears to be present also in the spectrum of \cite{Schonell17}, but not mentioned by the authors, whereas the latter unidentified line is not seen in their spectrum. No transitions near these two wavelengths are reported in observations of classical novae (\citealt{Wagner96}). We fit these unknown features with narrow Gaussians, but the $1.2612$~\um\ line is very weak and we could only obtain an upper limit on its flux. Therefore, we have not considered this line further in the fits of the individual spectra. In order to not make any prior assumptions about the \sn\ line profile, the narrow window containing the \sn\ line was masked and not used in the fitting process. 
Having fit for \pab, \feii~1.2570~\um\ and the unidentified line at $1.2483$~\um, we take the residual flux in the emission line complex as the \sn\ profile (Fig. \ref{fig:mean_spec_fits}).

\subsubsection{The \silvi~line} \label{sec:mean_si_vi}
The \silvi\ coronal line is blended with both the red wing of the weak hydrogen \brd\ broad line and the extreme red wing of the hydrogen \paa\ broad line. A very weak \htwo\ line is also expected at 1.9570~\um, although we did not convincingly detect it. A narrow window containing the \silvi\ line was masked and all other emission lines were modelled with Gaussians.

We assumed that the broad components of the hydrogen lines \paa\ and \brd\ have the same double-Gaussian profile as the \pab\ broad component (see Section~\ref{sec:mean_s_ix}). The amplitude of the broad \brd\ profile was fixed so as not to exceed the observed flux. Since the width of the narrow \brd\ line hit the maximum allowed value, we fixed it to that of the narrow \pab\ line and fit only for the flux and velocity offset. The resulting ratio between the \brd\ and \pab\ narrow line fluxes is $0.14\pm0.03$, which is similar to the value of 0.11 expected for Case B and a gas temperature of 15000~K (\citealt{OF06}). We note that an additional, weak broad Gaussian component was required by the fits in order to adequately model the red wing of \paa, but we do not ascribe a particular physical meaning to this component. 
The residuals of our best-fit model reveal the profile of the relatively strong \silvi\ line. It shows a prominent and extended red wing and excess blue flux in addition to a narrow core (Figs.~\ref{fig:mean_spec_fits} and \ref{fig:coronal_line_Gauss}).
\textcolor{black}{The width, flux and velocity offset of the core component have been determined by fitting a Gaussian to a region spanning $\approx800$~\kms\ across the peak of the line (the solid red line in Fig.~\ref{fig:coronal_line_Gauss}); these are the values reported for the `core' component in Table~\ref{tab:mean_fits}.
To determine the fluxes in the wings, we have integrated the excess flux from the peak of the core out to $\approx-2000$~\kms\ on the blue side of the core and $\approx+3000$~\kms\ on the red side.
The integration limits for the wings were determined by a visual inspection of the spectrum to determine the maximum velocity extents of the excess flux.
Since the red wing in particular is shallow at its extremity, it is likely that an appreciable amount of the integrated flux is just noise.
To account for this, we estimate the noise contribution to the integrated flux.
We first calculate the noise in the pseudo-continuum in two windows adjacent to the wings.
Using this value, we then simulate 1000 Gaussian noise `spectra' on the same velocity grid over which the wings are integrated.
The positive fluxes in these mock spectra are integrated and the mean value gives us an estimate of the amount of noise included in the wings' flux, which we take to be the uncertainty.
These values are reported for the `blue wing' and `red wing' components in in Table~\ref{tab:mean_fits}.}

\subsubsection{The \se~line}
\label{sec:mean_s_viii}
The \se\ coronal line is located near the emission maximum of a blend between the hydrogen \pad\ and \heii\ broad emission lines and close to their respective narrow components. This complex also contains the narrow, forbidden lines \ci\ 0.9827 and 0.9853~\um\ and \sii\ 1.0290, 1.0323, 1.0339 and 1.0373~\um. We fit the wavelength region of 0.975--1.027~\um, which excludes the \sii\ lines. We again used the broad \pab\ profile as a template for the broad hydrogen lines (here \pad) and we fit for its flux only. On the red side of the emission line blend, a single broad Gaussian was adequate to fit the remaining flux from broad \heii. Single Gaussians were included to model the \ci, \pad\ and \heii\ narrow lines. The velocity offsets of the two \ci\ lines were tied together.

Having modelled and subtracted emission lines other than \se\ in the complex, we take the residual flux from our model as the \se\ profile (see Fig.~\ref{fig:mean_spec_fits}). 
The \se\ coronal line has a similar profile to that of \silvi\, comprising of a red wing, excess blue flux and a narrow core (Figs.~\ref{fig:mean_spec_fits} and \ref{fig:coronal_line_Gauss}).
\textcolor{black}{Measurements of the line core and the blue and red wings have been made in the same manner as for \silvi, described in Section~\ref{sec:mean_si_vi}; the results are reported in Table~\ref{tab:mean_fits}.}

\subsubsection{Other emission lines}
For our study, we have modelled also the hydrogen \brg\ broad emission line, since it is of interest for black hole mass determinations in AGN (see Section~\ref{sec:mass}), and the \siii~$\lambda\lambda9069,9531$ narrow emission line doublet, since these lines can inform us about the intrinsic profile of transitions from the narrow emission line region.

There are no other broad emission lines in the vicinity of \brg\ and we isolated the line from a linear local continuum. The broad component was modelled with the same profile as the other hydrogen lines and we fit only for its flux. In addition to a narrow Gaussian for \brg, a second narrow Gaussian was added to model the \htwo~2.0332~\um\ emission line, which sits on the blue wing of broad \brg.

The \siii~$\lambda\lambda9069,9531$ narrow emission line doublet is a strong feature in our spectra. We subtracted the hydrogen \pae\ broad emission line from beneath the \siii\ lines, again using broad \pab\ as a template. We also subtracted a narrow Gaussian line at the rest-frame wavelength of \pae\ from the \siii~$\lambda9531$ line profile. 
Having accounted for the \pae\ emission, each \siii\ line was adequately fit with a single Gaussian of $\mathrm{FWHM}\approx370$~\kms.
No significant velocity shift was measured between the two \siii\ lines.
The centroid velocity shifts of all other lines measured in the mean spectrum were measured relative to \siii~$\lambda9531$.
The observed flux ratio of \siii~$\lambda9531$/\siii~$\lambda9069=2.5\pm0.1$ is consistent with the theoretical value of 2.58.

\subsection{The single-epoch spectra} \label{singlespec}
Because of the lower quality of the single-epoch spectra relative to the mean spectrum, multi-Gaussian fits to the emission line complexes were not always successful in reliably isolating the coronal lines. Therefore, we used their emission-line profiles determined in the mean spectrum as a guide. Having subtracted the underlying broad-band continuum following the method described in Section~\ref{sec:continuum} we then performed a fit to the emission line blend.
The mean \pab\ profile was again used as a template for all of the hydrogen lines; the template was rescaled in flux to match the hydrogen lines in the single-epoch spectra.
Each narrow line profile was modelled as a single Gaussian with its parameters taken from the fit to the same line in the mean spectrum.
If necessary, small scaling adjustments were made to the narrow lines to improve the fit.
\textcolor{black}{To some spectra we added a local continuum to correct cases in which we judged the broad-band continuum placement could be improved.}
The \textcolor{black}{local} continuum and contaminating lines \textcolor{black}{were} adjusted to obtain a good fit to the data then subtracted from the spectrum, leaving just the coronal line profile. 
\textcolor{black}{As can be seen in Figs.~\ref{fig:siii_variations}--\ref{fig:s_ix_variations} the residuals away from the coronal line are generally featureless (with the exception of the wings near the \se\ and \silvi\ profiles, and some artefacts of imperfect telluric correction around \sit) indicating that this procedure generally worked well.}
\textcolor{black}{However, t}he coronal line profiles in a few epochs were of very poor quality and we have excluded them for our further studies. 
\textcolor{black}{As for the mean spectrum, the fluxes, widths and velocity offsets of the narrow line cores are determined by fitting a Gaussian to the isolated emission line profile.
Again, for \se\ and \silvi\ we fit only the central portion of the profiles (within a few hundred \kms\ of the peak) to avoid the wings on either side of the line.}

\textcolor{black}{Because the coronal lines are low-contrast features in our spectra the FWHM and flux measurements are very sensitive to the placement of the underlying pseudo-continuum.
If the continuum is placed too high, for example, then both the FWHM and flux will be underestimated in the fitted profile.
This is an issue in the single-epoch spectra, in which the precise continuum placement is less certain than in the high-S/N mean spectrum.
We therefore calculate the range in both FWHM and flux that result from moving the the local continuum to its highest and lowest plausible levels (i.e.\ $\pm1\sigma$).
The value of $\sigma$, the noise in the continuum, is calculated from the standard deviation of flux points about the mean value in featureless windows near to each coronal line (and typically this noise was $\sim1$~per cent of the continuum flux). 
This additional uncertainty was added in quadrature to the Gaussian measurement uncertainties calculated by the fitting algorithm.
A similar approach was taken to estimate the uncertainties on the \se\ and \silvi\ wings; we calculate the error range from the maximum and minimum excess flux integrated with the continuum is moved $\pm1\sigma$.
Because the blue wing is weak and the red wing is very shallow at its extremity, this results in rather large uncertainties for the wings (particularly in the noisy part of the spectrum containing \se) and there are several epochs where no flux in excess of the Gaussian core is detected with confidence.}

Our results for the four near-IR coronal lines as well as the strong \siii~$\lambda9531$ line are shown in Figs. \ref{fig:siii_variations}-\ref{fig:s_ix_variations}\textcolor{black}{.  The fluxes of the line cores and wings over the course of the campaign} are recorded in Table~\ref{tab:core_wing_fluxes}.

\begin{table*}
    \centering
    \caption{Fluxes of the cores and blue and red wings in the profiles of \siii$\lambda9531$ and the near-infrared the coronal lines} 
    \begin{tabular}{l|ccccccccc}
    \hline
        & \siii\ & \multicolumn{3}{c}{\se} & \sn\ & \sit\ & \multicolumn{3}{c}{\silvi} \\
        &        & \multicolumn{3}{c}{$\overbrace{\hspace{4.5cm}}$} &  &  & \multicolumn{3}{c}{$\overbrace{\hspace{4.5cm}}$} \\
         Date & Core & Blue & Core & Red  & Core & Core & Blue & Core & Red \\
        &      & \textcolor{black}{wing} &      & \textcolor{black}{wing} &      &      & \textcolor{black}{wing} &      & \textcolor{black}{wing} \\
    \hline
    Mean spectrum & $33\pm3$ & $1.7\pm0.4$ & $7.3\pm0.4$  & $2.8\pm0.7$ & $4.3\pm0.2$ & $7.9\pm0.6$ & $3.6\pm0.4$ & $13.0\pm0.6$ & $8.7\pm0.6$   \\
    \hline
    \textcolor{black}{602}  \textcolor{black}{16/08/02} & $34\pm2$ & $3.5\pm\textcolor{black}{1.8}$ & $8.9\pm\textcolor{black}{3.4}$  & $4.4\pm\textcolor{black}{3.2}$ & $4.5\pm\textcolor{black}{1.3}$ & $7.7\pm\textcolor{black}{2.5}$ & $2.0\pm\textcolor{black}{1.1}$ & $13.2\pm\textcolor{black}{1.7}$                  & $18.7\pm\textcolor{black}{2.4}$  \\
    \textcolor{black}{611}  \textcolor{black}{16/08/11} & $35\pm2$ & $1.5\pm\textcolor{black}{1.9}$ & $7.2\pm\textcolor{black}{5.0}$  & $4.7\pm\textcolor{black}{2.9}$ & -                            & $5.7\pm\textcolor{black}{2.6}$ & $1.7\pm\textcolor{black}{1.3}$ & $12.4\pm\textcolor{black}{1.8}$                  & $16.7\pm\textcolor{black}{2.7}$  \\
    \textcolor{black}{743}  \textcolor{black}{16/12/21} & -        & -                            & -                             & -                            & $3.6\pm\textcolor{black}{1.3}$ & $6.0\pm\textcolor{black}{2.5}$ & -           & -                                              & - \\
    \textcolor{black}{759}  \textcolor{black}{17/01/06} & $34\pm2$ & $<\textcolor{black}{4.6}$      & $8.0\pm\textcolor{black}{3.7}$  & $7.0\pm\textcolor{black}{4.1}$ & $3.6\pm\textcolor{black}{1.8}$ & $7.9\pm\textcolor{black}{2.7}$ & $2.4\pm\textcolor{black}{0.5}$ & $13.3\pm\textcolor{black}{1.7}$                  & $<\textcolor{black}{2.1}$  \\
    \textcolor{black}{773}  \textcolor{black}{17/01/20} & $34\pm2$ & $4.1\pm\textcolor{black}{1.8}$ & $5.3\pm\textcolor{black}{3.4}$  & $<\textcolor{black}{6.2}$      & $5.2\pm\textcolor{black}{1.4}$ & $7.2\pm\textcolor{black}{2.7}$ & $3.9\pm\textcolor{black}{0.7}$ & $11.3\pm\textcolor{black}{1.7}$                  & $3.4\pm\textcolor{black}{1.8}$   \\
    \textcolor{black}{777}  \textcolor{black}{17/01/24} & $33\pm2$ & -                            & -                             & -                            & -                            & $6.5\pm\textcolor{black}{3.6}$ & $4.7\pm\textcolor{black}{1.1}$ & $1\textcolor{black}{4.7}\pm\textcolor{black}{2.3}$ & $5.5\pm\textcolor{black}{2.4}$   \\
    \textcolor{black}{789}  \textcolor{black}{17/02/05} & $33\pm1$ & $3.4\pm\textcolor{black}{2.2}$ & $6.6\pm\textcolor{black}{3.3}$  & $\textcolor{black}{<7.2}$      & $6.1\pm\textcolor{black}{1.7}$ & $8.1\pm\textcolor{black}{2.6}$ & $4.3\pm\textcolor{black}{0.8}$ & $13.9\pm\textcolor{black}{1.7}$                  & $3.4\pm\textcolor{black}{1.8}$   \\
    \textcolor{black}{799}  \textcolor{black}{17/02/15} & $35\pm2$ & $<\textcolor{black}{3.8}$      & $4.9\pm\textcolor{black}{3.3}$  & $3.6\pm\textcolor{black}{3.5}$ & $2.5\pm\textcolor{black}{1.3}$ & $6.6\pm\textcolor{black}{2.6}$ & $4.3\pm\textcolor{black}{0.9}$ & $14.7\pm\textcolor{black}{1.9}$                  & $4.9\pm\textcolor{black}{2.0}$   \\
    \textcolor{black}{804}  \textcolor{black}{17/02/24} & $34\pm1$ & $2.4\pm\textcolor{black}{1.7}$ & $4.5\pm\textcolor{black}{3.3}$  & $3.4\pm\textcolor{black}{3.4}$ & $4.1\pm\textcolor{black}{1.5}$ & $5.5\pm\textcolor{black}{2.6}$ & $2.8\pm\textcolor{black}{1.1}$ & $11.5\pm\textcolor{black}{1.8}$                  & $3.8\pm\textcolor{black}{2.4}$   \\
    \textcolor{black}{829}  \textcolor{black}{17/03/17} & $35\pm2$ & $2.9\pm\textcolor{black}{1.9}$ & $5.3\pm\textcolor{black}{3.2}$  & $3.7\pm\textcolor{black}{3.3}$ & $4.2\pm\textcolor{black}{1.7}$ & $5.8\pm\textcolor{black}{2.6}$ & $2.3\pm\textcolor{black}{0.7}$ & $11.3\pm\textcolor{black}{1.9}$                  & $2.2\pm\textcolor{black}{1.7}$   \\
    \textcolor{black}{834}  \textcolor{black}{17/03/22} & $35\pm2$ & $<\textcolor{black}{2.7}$      & $6.3\pm\textcolor{black}{3.3}$  & $5.3\pm\textcolor{black}{2.4}$ & $5.5\pm\textcolor{black}{1.5}$ & $8.2\pm\textcolor{black}{3.0}$ & $2.3\pm\textcolor{black}{1.1}$ & $15.3\pm\textcolor{black}{2.0}$                  & $\textcolor{black}{<5.1}$   \\
    \textcolor{black}{882}  \textcolor{black}{17/05/09} & $33\pm2$ & -                            & -                             & -                            & $3.8\pm\textcolor{black}{1.8}$ & $6.7\pm\textcolor{black}{2.6}$ & $3.1\pm\textcolor{black}{0.7}$ & $12.0\pm\textcolor{black}{1.7}$                  & $4.1\pm\textcolor{black}{1.5}$   \\
    \textcolor{black}{898}  \textcolor{black}{17/05/25} & $33\pm3$ & $4.0\pm\textcolor{black}{2.7}$ & $8.9\pm\textcolor{black}{4.7}$  & $7.0\pm\textcolor{black}{4.5}$ & $5.2\pm\textcolor{black}{1.7}$ & $9.1\pm\textcolor{black}{3.1}$ & $4.0\pm\textcolor{black}{1.1}$ & $15.3\pm\textcolor{black}{1.7}$                  & $4.9\pm\textcolor{black}{2.4}$   \\
    \textcolor{black}{914}  \textcolor{black}{17/06/10} & $34\pm2$ & -                            & -                             & -                            & -                            & -                            & $3.7\pm\textcolor{black}{0.8}$ & $13.2\pm\textcolor{black}{1.8}$                  & $6.7\pm\textcolor{black}{1.7}$   \\
    \textcolor{black}{924}  \textcolor{black}{17/06/20} & $33\pm2$ & $\textcolor{black}{<6.4}$      & $12.5\pm\textcolor{black}{6.9}$ & $6\textcolor{black}{.3\pm5.0}$ & $3.2\pm\textcolor{black}{1.9}$ & $7.3\pm\textcolor{black}{4.5}$ & $4.8\pm\textcolor{black}{1.3}$ & $10.5\pm\textcolor{black}{1.7}$                  & $9.9\pm\textcolor{black}{2.9}$   \\
    \textcolor{black}{932}  \textcolor{black}{17/06/28} & $33\pm2$ & $3.0\pm\textcolor{black}{1.5}$ & $7.8\pm\textcolor{black}{3.3}$  & $7.7\pm\textcolor{black}{2.8}$ & $1.6\pm\textcolor{black}{1.2}$ & $7.0\pm\textcolor{black}{3.4}$ & $5.6\pm\textcolor{black}{1.1}$ & $9.7\pm\textcolor{black}{1.8}$                   & $9.3\pm\textcolor{black}{2.5}$   \\
    \textcolor{black}{937}  \textcolor{black}{17/07/03} & $33\pm2$ & $2.7\pm\textcolor{black}{1.6}$ & $9.4\pm\textcolor{black}{3.4}$  & $5.3\pm\textcolor{black}{2.9}$ & -                            & -                            & -           & -            & -     \\
    \hline
    RMS         & 0.84 (2.5\%) & 0.\textcolor{black}{62} (\textcolor{black}{20}\%) & 2.\textcolor{black}{25} (3\textcolor{black}{2}\%)  & 1.5\textcolor{black}{0} (\textcolor{black}{29}\%) & 1.25 (3\textcolor{black}{1}\%) & 1.0\textcolor{black}{3} (15\%) & 1.18 (3\textcolor{black}{5}\%) & 1.\textcolor{black}{69} (13\%)  & 5.\textcolor{black}{22} (\textcolor{black}{79}\%)      \\   
    \hline
    \end{tabular}
    \parbox[]{0.95\textwidth}{\textcolor{black}{The observation date is given in the formats $\mathrm{MJD}-57000$ and YY/MM/DD and the fluxes are in units $10^{-15}$~erg\,s$^{-1}$\,cm$^{-2}$.}}
    \label{tab:core_wing_fluxes}
\end{table*}

\subsection{Optical iron coronal lines}
\label{sec:optical}

During one of our other observing programs, we obtained a high-resolution optical spectrum of \ngc\ in 2015 March at the William Herschel Telescope (WHT), a 4~m telescope on La Palma.
We dereddened this spectrum using $E(B-V)=0.0168$ (\citealt{SF11}) and the extinction curve of \cite{CCM89}.
The spectrum contains six high-ionisation forbidden lines of iron:
[Fe\,\textsc{vii}]$\lambda3759,5159,5721,6087$, [Fe\,\textsc{x}]$\lambda6374$ and [Fe\,\textsc{xi}]$\lambda7892$.
With the exception of [Fe\,\textsc{x}] these lines are unblended so we simply measure their profiles with a single Gaussian plus linear local continuum.  
[Fe\,\textsc{x}] is blended with [O\,\textsc{i}]$\lambda6363$ so we first determine the profile of  [O\,\textsc{i}]$\lambda6300$ and use this as a template for [O\,\textsc{i}]$\lambda6363$, assuming a 1:3 flux ratio. 
We assume \oiii$\lambda5007$ to be at rest within \ngc\ and measure all other line velocity shifts relative to it.
The FWHMs of the lines are all corrected for instrumental broadening, assuming this is $\sim250$~km\,s$^{-1}$ for the ISIS instrument on the WHT.
The properties of these lines are reported in Table~\ref{tab:opt_fe}.

\begin{table*}
    \centering
    \caption{Properties of optical iron coronal lines and measurements from the 2015 WHT spectrum}
    \label{tab:opt_fe}
    \begin{tabular}{ccccccc}
    \hline
    Ion                  & $\lambda$ & $\chi$ & $\log(n_\mathrm{crit})$ & Velocity offset & FWHM          & Flux \\
                         & [\AA]     & [eV]   & [cm$^{-3}$]              & [km\,s$^{-1}$]   & [km\,s$^{-1}$] & [$10^{-15}$ erg\,s$^{-1}$\,cm$^{-2}$] \\
    (1)                  & (2)       & (3)    & (4)                     & (5)             & (6)           & (7) \\
    \hline
    {[Fe\,\textsc{vii}]} & 3759      & 99.1   & 7.60                    & $-207\pm17$     & $559\pm38$    & $16.9\pm1.5$ \\
    "                    & 5159      & 99.1   & 6.54                    & $-126\pm21$     & $<250$        & $3.01\pm0.80$ \\
    "                    & 5721      & 99.1   & 7.57                    & $-111\pm11$     & $485\pm18$    & $10.5\pm0.5$ \\
    "                    & 6087      & 99.1   & 7.64                    & $-201\pm9$      & $632\pm13$    & $20.1\pm0.5$ \\
    {[Fe\,\textsc{x}]}   & 6374      & 233.6  & 8.64                    & $-188\pm15$     & $605\pm46$    & $5.24\pm0.51$ \\
    {[Fe\,\textsc{xi}]}  & 7892      & 262.1  & 8.81                    & $-212\pm24$     & $579\pm57$    & $4.44\pm0.56$ \\
    \hline
    \end{tabular}
\parbox[]{11cm}{The columns are: (1) Ion; (2) vacuum wavelength; (3) ionisation potential; (4) critical density for $\log(T_\mathrm{e}/\mathrm{K})=4$; (5) velocity offset of the emission line peak relative to \oiii$\lambda5007$~\AA; (6) full width at half maximum of the emission line component corrected for an instrumental broadening of 250~\kms; and (7) line flux. We give $1\sigma$ errors.}
\end{table*}

\section{Results} \label{results}
\subsection{The \textcolor{black}{mean} coronal line profiles}
\label{sec:profiles}
Fig. \ref{fig:coronal_line_Gauss} shows the profiles of the four near-IR coronal lines as observed in the mean spectrum. In all four cases, the central (core) part can be modelled well with a single Gaussian. It is intriguing to notice that both the \se\ and \silvi\ coronal lines show significant excess flux blue- and red-ward of the core, whereas the \sn\ and \sit\ coronal lines do not. 

\begin{figure}
    \centering
    \includegraphics[width=1\columnwidth]{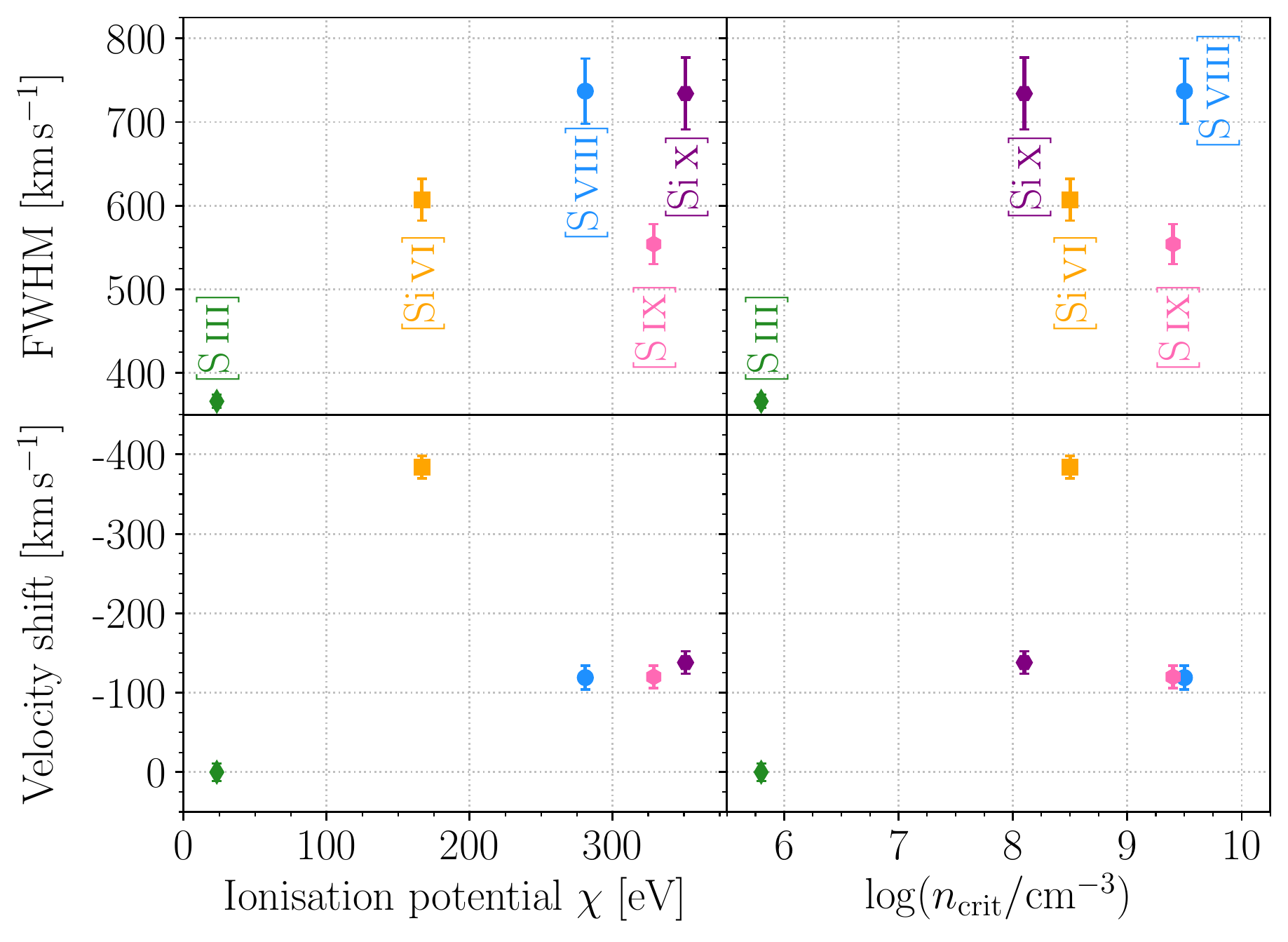}
    \caption{FWHM (top) and centroid velocity shift (bottom) of the coronal line and \siii$\lambda9531$ cores, measured in the mean spectrum, as a function of ionisation potential (left) and critical density (right).}
    \label{fig:ip}
\end{figure}

In Fig.~\ref{fig:ip} we show the coronal line and \siii\ widths and velocity shifts as a function of both their ionisation potentials and critical densities.
The coronal line (core) profiles are all broader than the \siii\ doublet lines, with line widths of $\mathrm{FWHM}\sim500$--$800$~\kms, compared with $\approx370$~\kms\ for \siii.
Whilst we have only five data points, Fig.~\ref{fig:ip} shows a trend of increasing FWHM with ionisation potential up to $\approx300$~eV.
It is clear in Fig.~\ref{fig:ip} that all four coronal lines have emission line peaks blue-shifted with respect to \siii.
The \se, \sn\ and \sit\ lines have very similar average velocity offsets of $\approx-130$~\kms; curiously, the lowest-ionisation potential coronal line \silvi\ has a much greater velocity offset of $\approx-380$~\kms, so there is no simple relationship between the velocity shift and ionisation potential. 
We also measured the velocity shifts of the other forbidden, narrow emission lines in the mean spectrum, with none of them having a significant velocity shift (see Table~\ref{tab:mean_fits}).
Although it is clear that the coronal lines have higher critical densities and both greater line widths and velocity shifts than \siii, no simple trend with critical density is observed between the four coronal lines. 

\subsection{The coronal line variability}
\label{sec:variability}
\begin{figure*}
	\includegraphics[width=1.8\columnwidth]{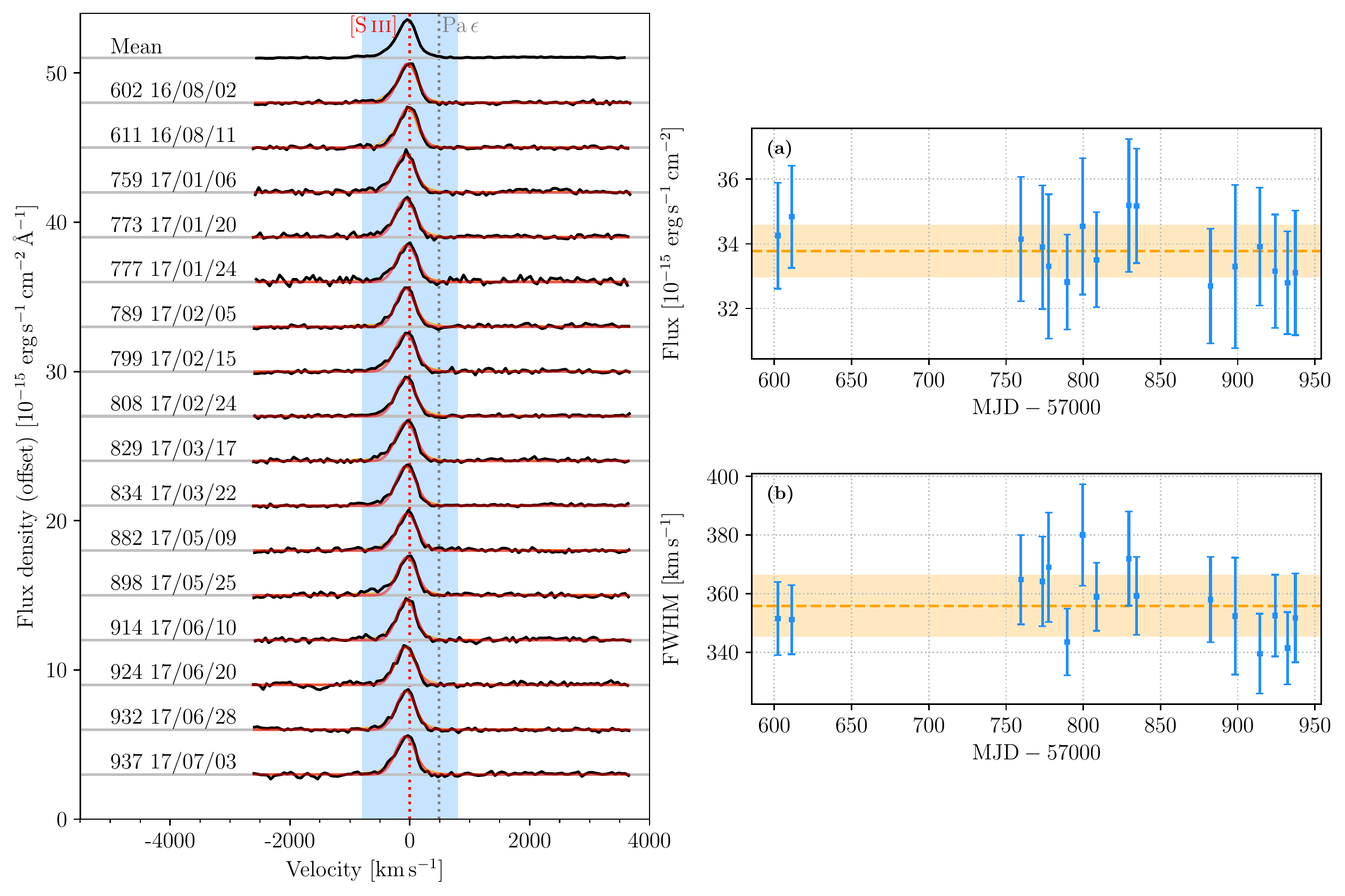}
    \caption{The \siii~$\lambda9531$ emission line in the mean and single-epoch spectra. 
    Left-hand panel: \siii\ emission line profiles isolated from other emission lines and the underlying continuum (black). The location of the subtracted contaminating emission lines is indicated with a grey dashed line. The \siii\ profile extracted from the mean spectrum is shown in the individual spectra by the thin orange line. In each spectrum the blue shaded region was fitted with a single Gaussian (red line).
    \textcolor{black}{Each spectrum is labelled with the observation date in the formats $\mathrm{MJD}-57000$ and YY/MM/DD.}
    Right-hand panels (a) and (b) show \textcolor{black}{v}ariations of the line core flux and FWHM obtained from the Gaussian fit. 
    The error-weighted mean flux and FWHM are shown by the orange dashed lines, and the orange shaded regions are the mean $\pm\sigma$.
    We plot $1\sigma$ errors.}
    \label{fig:siii_variations}
\end{figure*}

The low-ionisation narrow forbidden  emission lines are assumed to be produced in AGN at distances from the central engine large enough to not show significant flux variability over the course of decades. But the optical narrow-line reverberation results for \ngc\ show that the \oiii~$\lambda5007$ line emitting region is more nuclear than expected, with an extent of only 1--3~pc and a density of $\sim10^5$~cm$^{-3}$ \citep{Pet13}. Since our campaign extends over the course of roughly a year, we can still assume that the  flux of the \siii~$\lambda9531$ line is not variable. This expectation is confirmed by the results in Table~\ref{tab:core_wing_fluxes} and Fig.~\ref{fig:siii_variations}:
the RMS variability of \siii$\lambda9531$ is $\approx8.4\times10^{-16}$~erg\,s$^{-1}$\,cm$^{-2}$, only around 2.5~per~cent of the mean line flux.
The width of this line is also stable during our campaign, with two-thirds of data points being consistent with the mean value to within 1$\sigma$.

\subsubsection{The \silvi\ and \se\ lines}
\begin{figure*}
	\includegraphics[width=1.8\columnwidth]{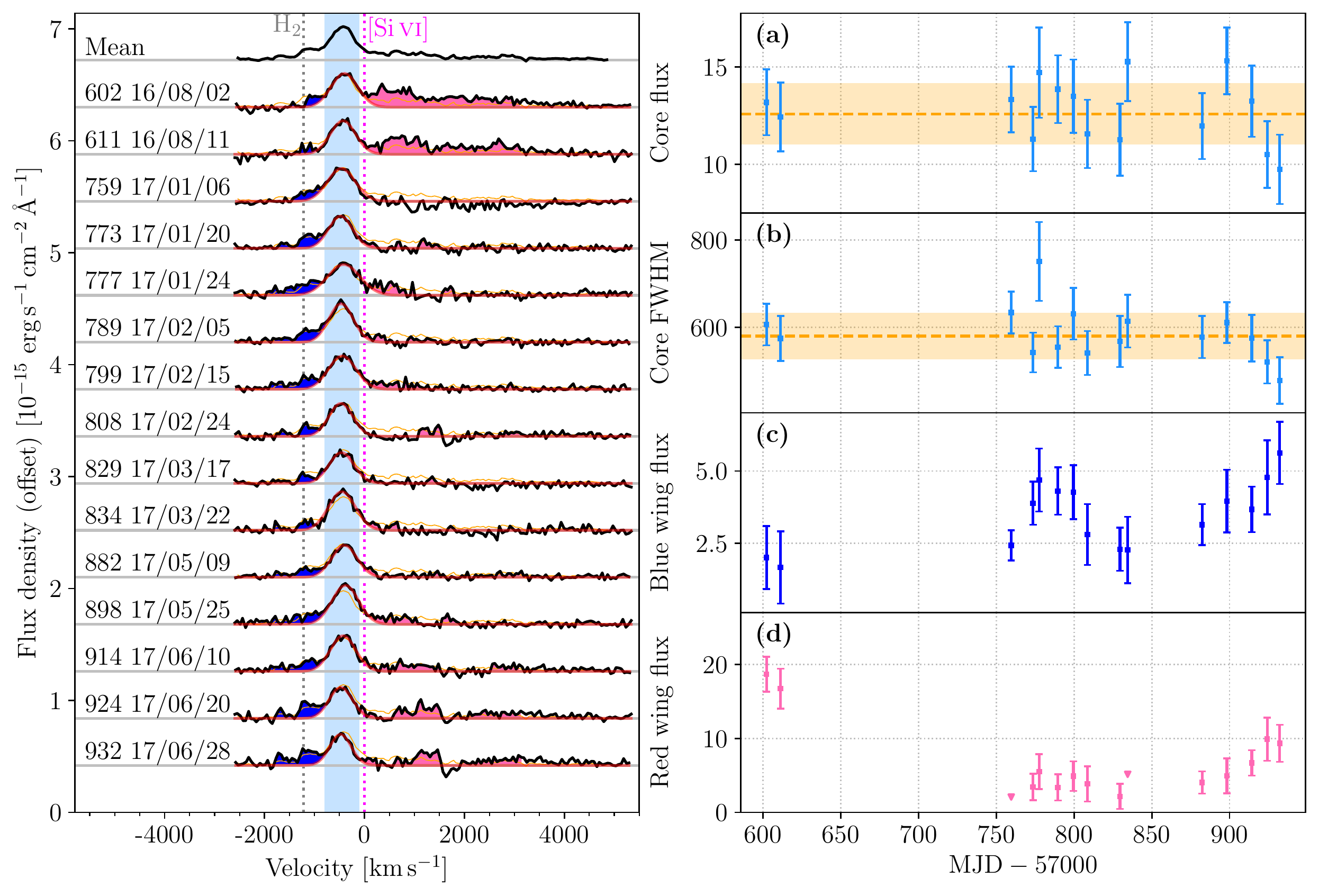}
    \caption{The \silvi\ coronal line in the mean and single-epoch spectra.     Left-hand panel: \silvi\ emission line profiles isolated from other emission lines and the underlying continuum (black). The locations of the contaminating emission lines are indicated with grey dashed lines; the rest-frame velocity of the coronal line is indicated by the magenta dotted line. 
    The \silvi\ profile extracted from the mean spectrum is shown in the individual spectra by the thin orange line.
    In each spectrum the light blue shaded region was fitted with a single Gaussian (red line).
    \textcolor{black}{Each spectrum is labelled with the observation date in the formats $\mathrm{MJD}-57000$ and YY/MM/DD.}
    Right-hand panels (a) and (b) show the flux and FWHM (in \kms) of this Gaussian.    
    The error-weighted mean flux is shown by the orange dashed line, and the orange shaded region spans the mean $\pm\sigma$.
    Panels (c) and (d) show the excess flux above the fitted Gaussian on the blue and red sides of the line centre, respectively.
    The excess fluxes which were integrated are shaded blue and pink in the left-hand panel.
    All fluxes are in units $10^{-15}$~erg\,s$^{-1}$\,cm$^{-2}$.
    We plot $1\sigma$ errors.}
    \label{fig:si_vi_variations}
\end{figure*}

The \silvi\ line is the strongest of the four near-IR coronal lines in our spectra;
we reliably isolated its profile in 15/18 spectra (Fig.~\ref{fig:si_vi_variations}). We measure a maximum flux variability of $\sim50$~per cent for the \silvi\ line core, with $\mathrm{RMS}\approx13$~per~cent  (Table \ref{tab:core_wing_fluxes}).
In addition, we observe variable excess emission blue- and red-ward of the line core. Both of these flux excesses are relatively broad, with the blue- and red-shifted part stretching over $\sim2000$ and $\sim3000$~\kms, respectively (Fig. \ref{fig:min_max_wings}), and in the mean spectrum their combined flux is similar to the flux in the core of the line. The red flux excess variability shows a cyclical behaviour: it is most prominent in the first spectra with a flux exceeding that in the line core, becomes weak by the middle of the campaign and increases in flux again at later epochs (Fig.~\ref{fig:si_vi_variations} and Table~\ref{tab:core_wing_fluxes}). 
In the second part of the campaign (after the seasonal gap) the blue excess varies in the same manner as the red excess however the blue excess is weakest in the first observing epochs, unlike the red excess, which is strongest.

\begin{figure*}
	\includegraphics[width=1.8\columnwidth]{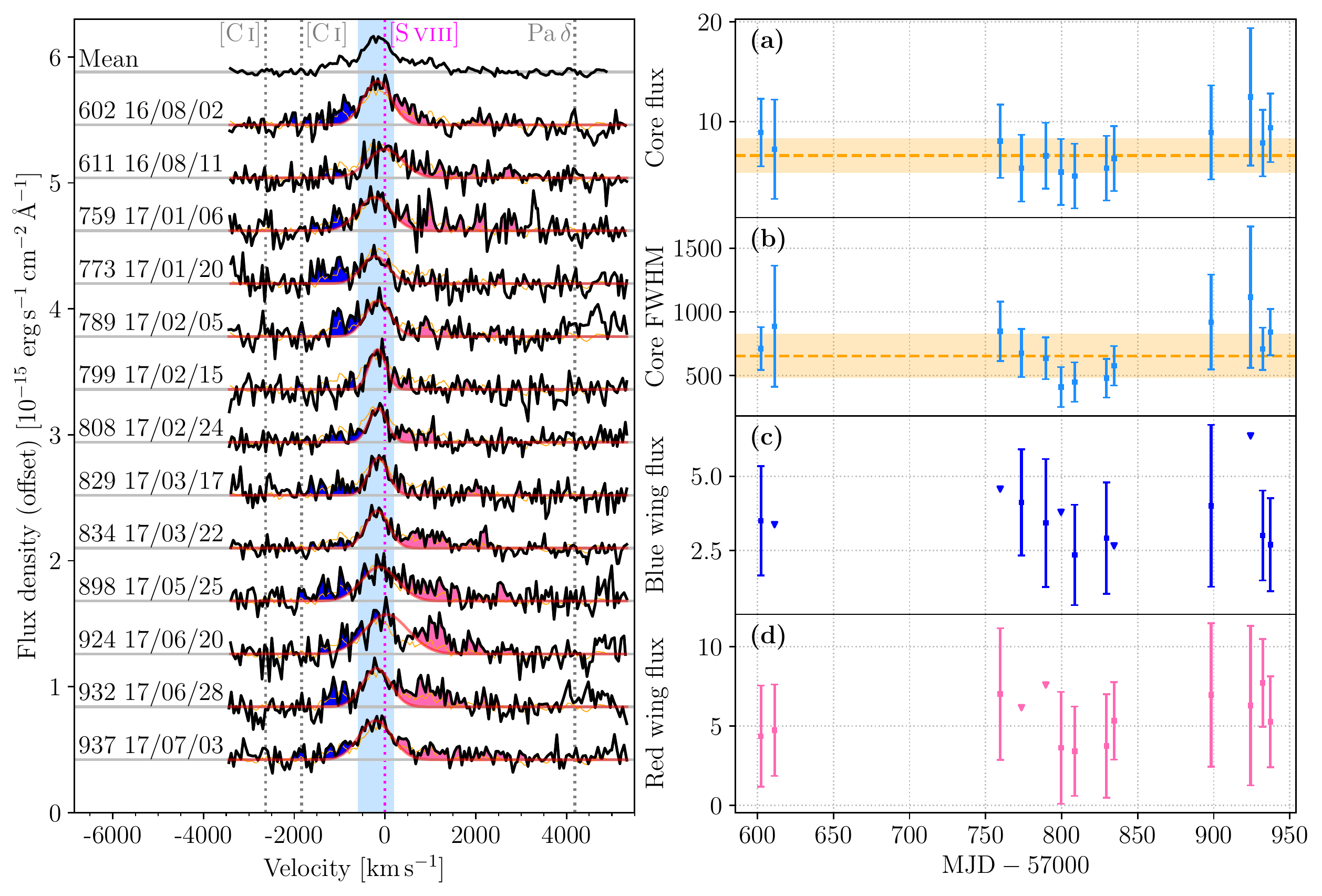}
    \caption{The \se\ coronal line in the mean and single-epoch spectra. Description as in Fig. \ref{fig:si_vi_variations}.}
    \label{fig:s_viii_variations}
\end{figure*}

\begin{figure*}
    \centering
    \begin{tabular}{cc}
        \includegraphics[width=.9\columnwidth]{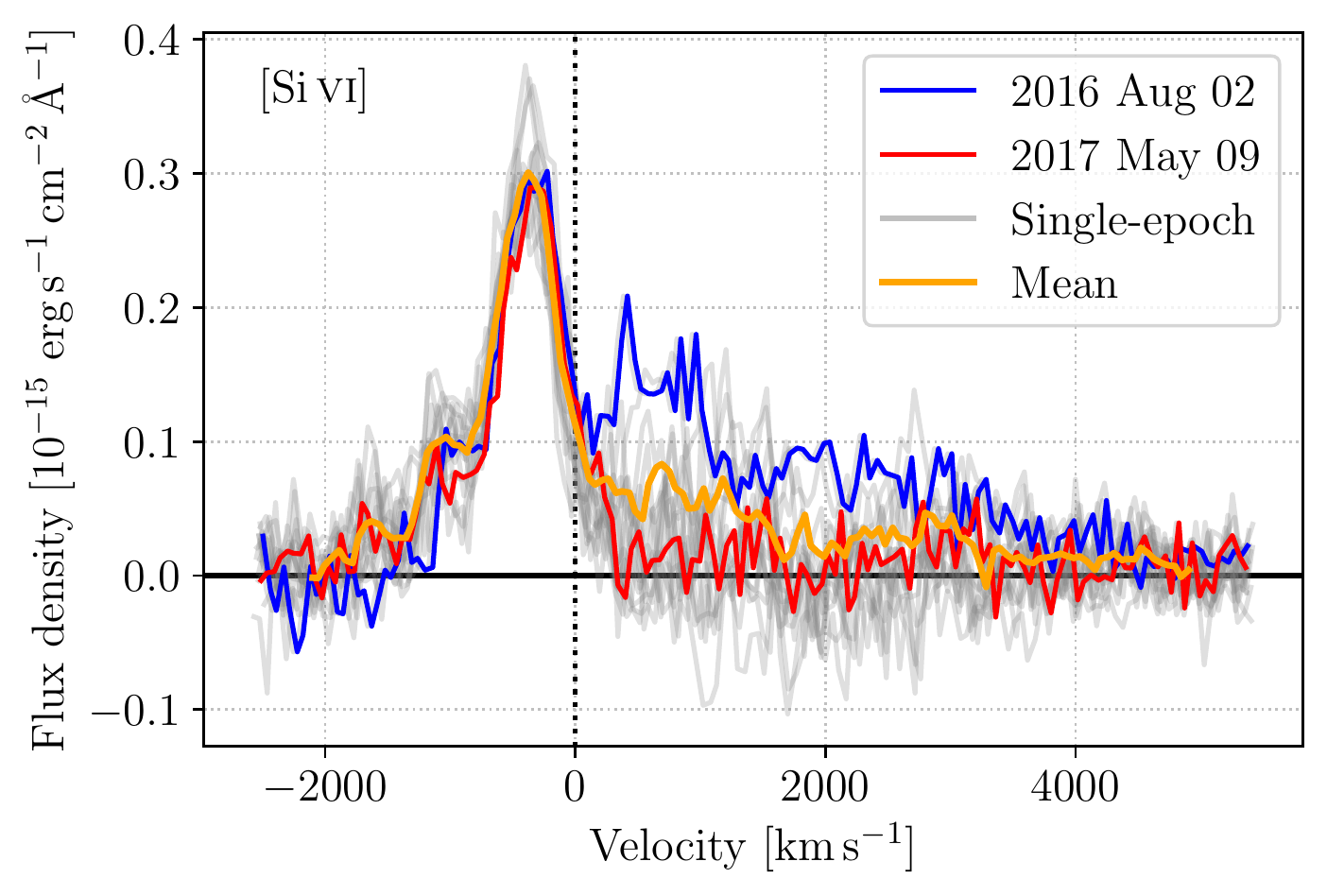} & \includegraphics[width=.9\columnwidth]{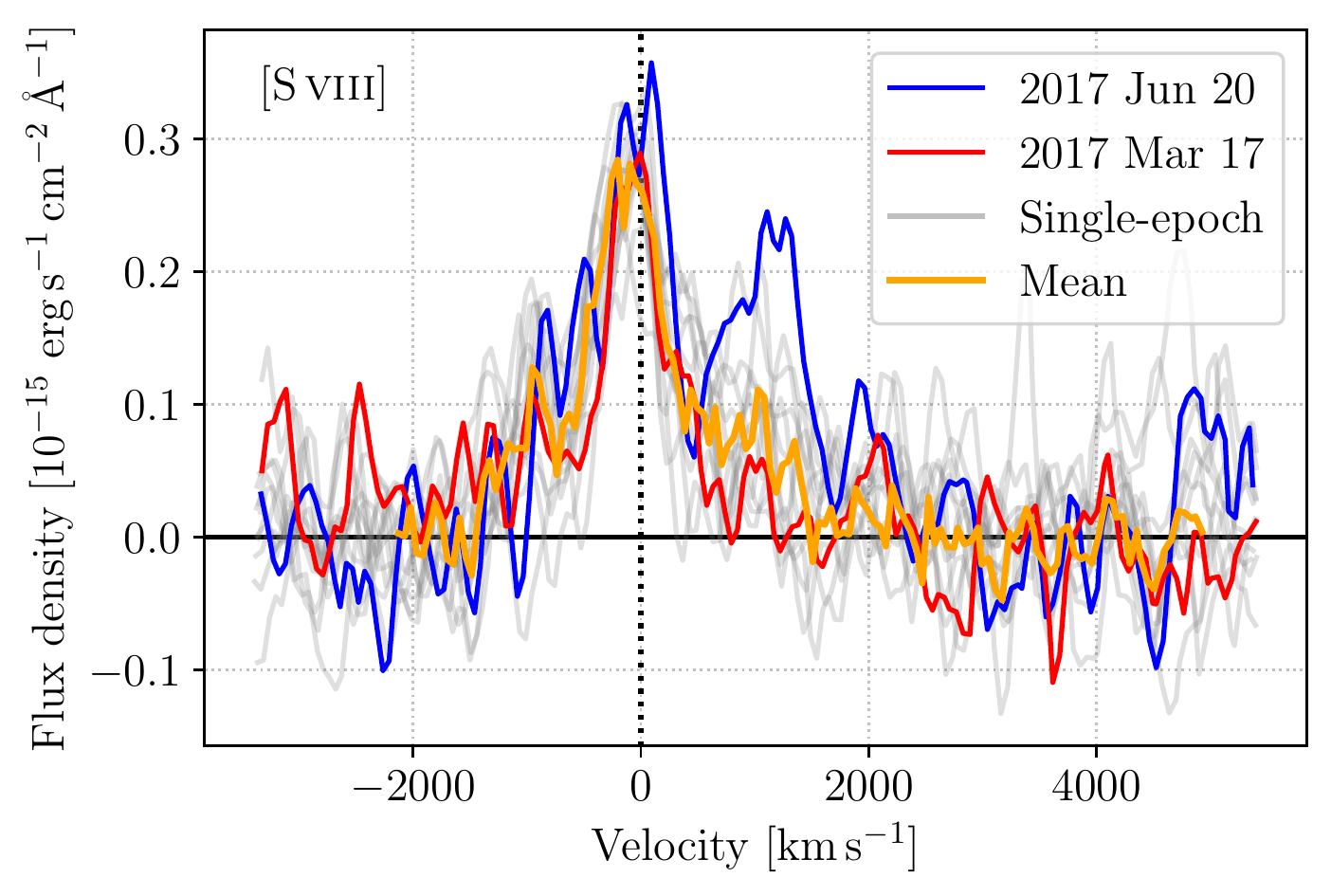} \\
    \end{tabular}
    \caption[]{The variability of the \silvi\ and \se\ coronal line profiles. 
    All extracted profiles are shown in grey and the mean line profiles in orange.
    An epoch with strong (weak) wings is highlighted in blue (red).
    The single-epoch \se\ profiles have been smoothed slightly for clarity.}
    \label{fig:min_max_wings}
\end{figure*}

The \se\ coronal line
was strong enough to be reliably isolated in 13/18 single-epoch spectra (Fig.~\ref{fig:s_viii_variations}). The \se\ line core is much more variable than that of the \silvi\ line, with maximum flux changes by a factor of $\approx3$ (Table~\ref{tab:core_wing_fluxes}). 
Like for the \silvi\ line, we observe variable and broad excess emission blue- and red-ward of the line core (Fig.~\ref{fig:min_max_wings}), but in the mean spectrum their combined flux is only about half that of the core of the line. Given that the \se\ line is also on average a factor of $\sim2$ weaker than the \silvi\ line, the isolation of the excess fluxes in the single-epoch spectra is more problematic. However, there is a trend for this excess emission to be strongly variable (by factors of a few) also in the \se\ coronal line.


In Fig.~\ref{fig:min_max_wings} we overplot all of the single-epoch line profiles of both \se\ and \silvi\ for easier comparison.
In \silvi\ it is clear that most of the variability in the profile is in the broad wings, with the narrow core showing relatively little variation from the mean.
This is reflected in the RMS variability measured for the different line components (Table~\ref{tab:core_wing_fluxes}).
Because of the noisier data, it is more difficult to see the same behaviour in the \se\ line and the RMS variability of the core and blue and red wings are similar ($\approx30$~per~cent).
In Fig.~\ref{fig:min_max_wings} we highlight two spectra with strong and weak wings but similar cores, so visually at least a similar trend as observed in \silvi\ can be discerned.

\subsubsection{The \sit\ and \sn\ lines}
\begin{figure*}
	\includegraphics[width=1.8\columnwidth]{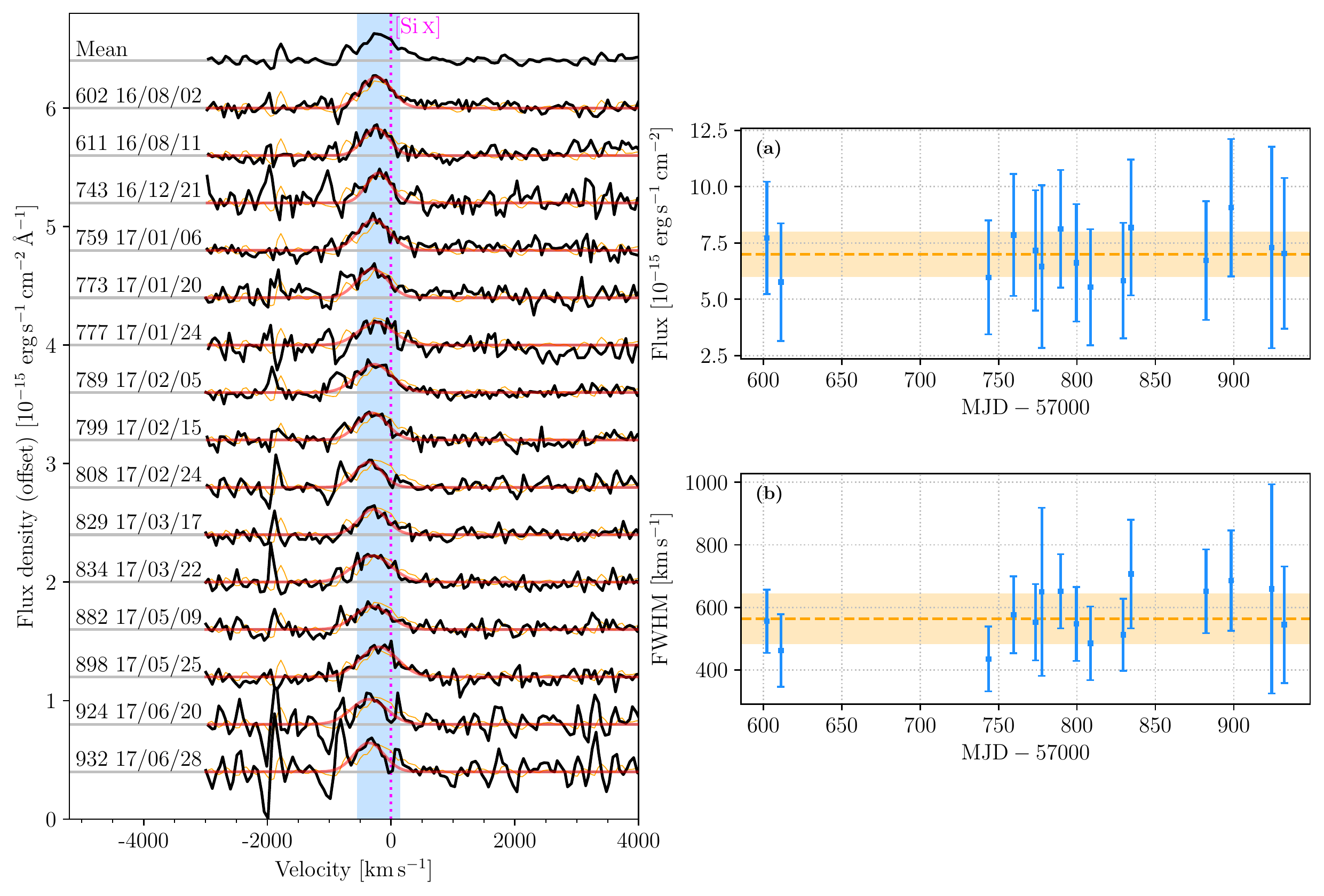}
    \caption{The \sit\ coronal line in the mean and single-epoch spectra. Description as in Fig. \ref{fig:siii_variations}.}
    \label{fig:si_x_variations}
\end{figure*}

The \sit\ line is the second strongest coronal line in our spectra and is free of blends. However, it is located in a region of telluric absorption, which is strongest blueward of it. 
We extracted its profile in 15/18 single-epoch spectra (Figure~\ref{fig:si_x_variations}).
\textcolor{black}{Its RMS flux variability (15~per cent) is very similar to that of the \silvi\ core.}

\begin{figure*}
	\includegraphics[width=1.8\columnwidth]{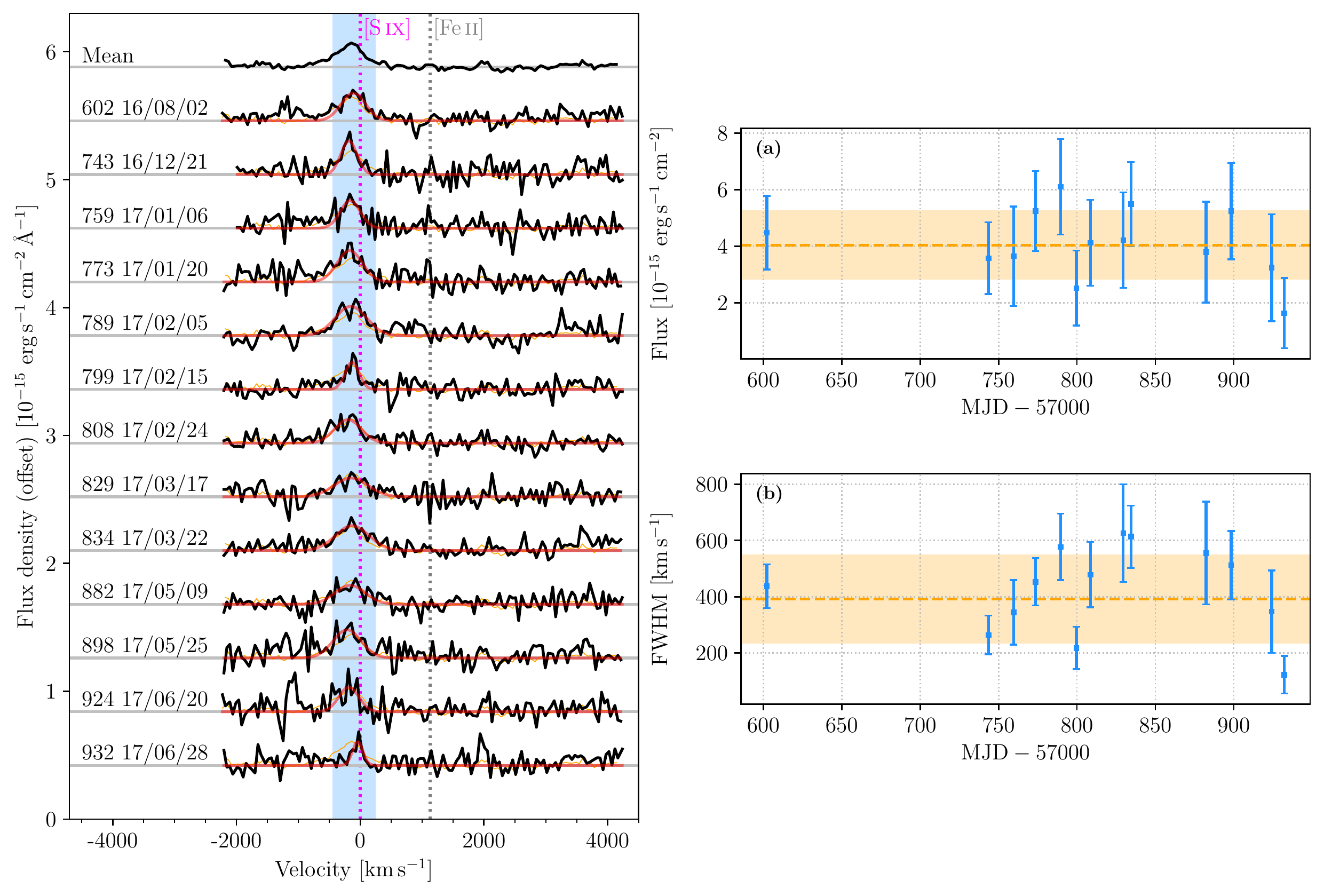}
    \caption{The \sn\ coronal line in the mean and single-epoch spectra. Description as in Fig. \ref{fig:siii_variations}.}
    \label{fig:s_ix_variations}
\end{figure*}

\sn\ is the weakest near-IR coronal line in our study but was strong enough to be reliably isolated in 13/18 single-epoch spectra. 
It shows a similar variability range to the \se\ line core, with maximum flux changes by a factor of $\sim3$ and RMS of 3\textcolor{black}{1}~per~cent, but without a clear trend (Fig.~\ref{fig:s_ix_variations}). 

\subsubsection{Trends in coronal line variability}
\label{sec:variability-trends}
\textcolor{black}{As can be seen from Table~\ref{tab:core_wing_fluxes}, the coronal line cores show a slightly higher degree of variability than that of \siii~$\lambda9531$: $\mathrm{RMS}\approx10$--30~per cent for the coronal lines compared with $\mathrm{RMS}=2.5$~per cent for \siii.
Figs.~\ref{fig:si_vi_variations}, \ref{fig:s_viii_variations}, \ref{fig:si_x_variations} and \ref{fig:s_ix_variations} suggest that there is a correspondence between the measured FWHM and flux of the coronal line cores, in the sense that as the flux increases the line becomes broader.
However, this variability is not statistically significant and simple fits for a constant line flux return $\chi^2_\nu=0.27$, 0.77, 0.14 and 0.85 for the \se, \sn, \sit\ and \silvi\ line cores, respectively.
Given the substantial uncertainties, we cannot make strong statements about the variability of the narrow line cores.}

\begin{figure}
    \centering
    \includegraphics[width=\columnwidth]{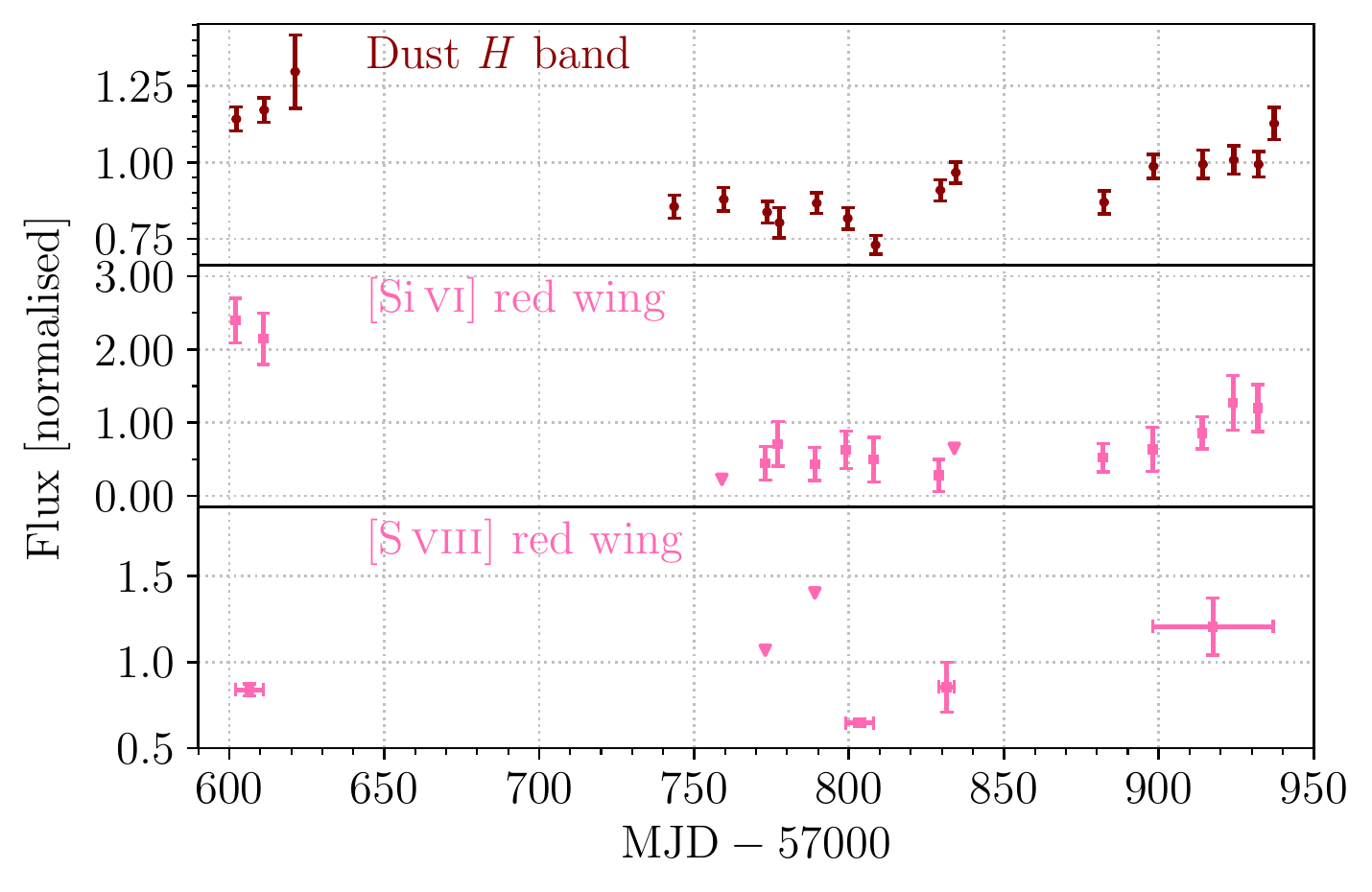}
    \caption{
    \textit{Top:} The \textit{H} band (1.55--1.60~\protect\um) dust flux, from \protect\cite{Landt19}.  
    \textcolor{black}{\textit{Middle and bottom:}} The \protect\silvi\ \textcolor{black}{and \protect\se} broad red wing flux\textcolor{black}{es}, from this work.
    \textcolor{black}{We have binned \protect\se\ measurements from consecutive epochs and plot the error-weighted averages and standard deviations. All fluxes are normalised to their mean value.}}
    \label{fig:dust_wings}
\end{figure}

\textcolor{black}{It is clear from Figs.~\ref{fig:si_vi_variations}, \ref{fig:s_viii_variations} and \ref{fig:min_max_wings} that most of the variability in the coronal line profiles is in the wings of \se\ and \silvi.
Flux variations in the \silvi\ red wing are statistically significant and we can reject the null hypothesis that the variability is purely statistical with greater than 99.99~per cent confidence.
A fit of constant flux to the \silvi\ blue wing lightcurve is poor ($\chi^2_\nu=1.45$), suggesting variability of the blue wing also, but the variability is less significant than in the red wing ($p\approx0.12$).}
In Fig.~\ref{fig:dust_wings} we show the variability of the \silvi\ \textcolor{black}{and \se} broad red wing\textcolor{black}{s} and compare them to that of the hot dust.
\textcolor{black}{Because of the noisier data, trends in the raw \se\ red wing lightcurve (Fig.~\ref{fig:s_viii_variations}d) are less clear.
We have therefore binned the lightcurve to better show the long-term trend and in Fig.~\ref{fig:dust_wings} we show the error-weighted average and standard deviations from the mean of consecutive flux points, as indicated.}
The dust lightcurve is derived from spectroscopic data, with fluxes integrated over a narrow emission line free region near the centre of the \textit{H} photometric band (1.55--1.60~\um; \citealt{Landt19}).
We observe a strong similarity in the shapes of the dust and the coronal line red wing lightcurves \textcolor{black}{(particularly \silvi)}: the fluxes are initially high, but weaken substantially during the gap in observations between MJD 57621 and 57743; the fluxes remain low in the middle of the campaign, before a systematic rise from MJD 57882 onwards.

\subsection{The black hole mass from near-IR emission line ratios}
\label{sec:mass}
\cite{Rod20} presented a novel method to calculate black hole masses using the flux ratio of coronal lines to low-ionisation permitted lines (see also \citealt{Prieto22}).
From flux measurements made by \cite{Riffel06}, \cite{Rod20} calculate the flux ratio \silvi/\brg$=(9.97\pm0.86)/(16.27\pm2.0)=0.61\pm0.09$
from which they determined a black hole mass of $6.7\times10^6$~M$_\odot$.
Considering the total flux (core $+$ wings) in the \silvi\ line, we calculate \silvi/\brg$=(25.3\pm0.9)/(43.3\pm0.3)=0.58\pm0.02$ from our mean spectrum,  
which is consistent within the uncertainties with the value determined by \cite{Rod20} and therefore gives the same black hole mass.
Therefore, although the absolute fluxes of the lines differ by more than a factor two between the 2002 IRTF spectrum of \cite{Riffel06} and our 2016--17 mean spectrum, the flux ratio has not changed, as one would expect if the flux ratio reflects the black hole mass.

We note that using the \silvi\ core flux only results in the flux ratio $=0.31\pm0.01$ and so the estimated black hole mass is approximately four times greater at $2.6\times10^7$~M$_\odot$, much closer to the $3.2\times10^{7}$~M$_\odot$ obtained by \cite{Pancoast14} via reverberation mapping.
The mass obtained from the total \silvi\ line flux is 0.66~dex discrepant with the \cite{Pancoast14} estimate, whereas the mass from the core flux is only 0.09~dex discrepant.
These compare with a 0.44~dex scatter in black hole mass reported by \cite{Rod20} for their scaling relation.

\section{Photoionisation models}
\label{sec:cloudy}

\begin{figure}
    \centering
    \includegraphics[width=\columnwidth]{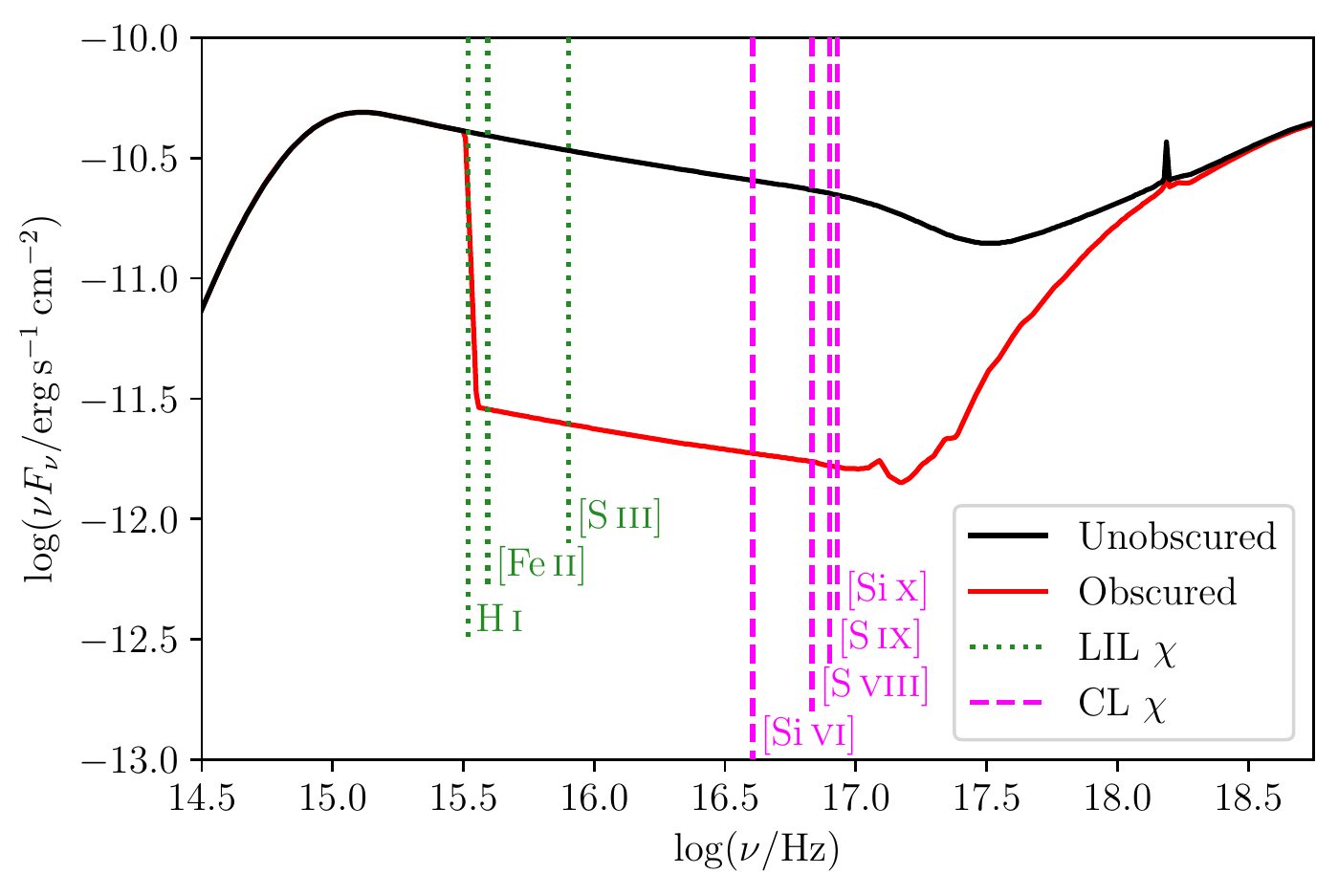}
    \caption{The unobscured and obscured SEDs of \ngc\ (\citealt{Mehdipour15}) are shown as solid black and red lines, respectively.  
    Dotted green lines indicate the ionisation potentials ($\chi$) of the low-ionisation lines (LIL) \hi, \feii\ and \siii.
    Dashed magenta lines show the ionisation potentials of the coronal lines (CL), ranging from 166.8~eV for \silvi\ to 351.1~eV for \sit\ (see Table~\ref{tab:contaminants}).}
    \label{fig:seds}
\end{figure}

AGN coronal lines are thought to be photoionised by the soft X-ray nuclear continuum.
Since ionic species are most effectively produced by photons with energies just above their ionisation potentials, the flux ratios of lines with different ionisation potentials will depend on the shape of the ionising SED.
The shape of the ionising SED depends not just on the properties of the accretion flow
(the black hole mass, accretion rate, electron temperature and optical depths of the warm and hot coronae, etc.) but also on extrinsic factors such as obscuration of the ionising source by gas and dust.
This is of particular relevance in studies of \ngc\ in which a persistent obscurer, first reported by \cite{Kaastra14}, strongly absorbs the nuclear soft X-ray emission along certain lines of sight.
\cite{Kaastra14} described the obscurer as a persistent, clumpy, ionised gas outflow on scales of only a few light-days from the nucleus.
\cite{Dehghanian19} interpreted it as the upper, line-of-sight component of an accretion disc wind; the wind originates interior to the BLR and its dense base shields the BLR from the nuclear continuum source.
We know that this obscurer was present during the 2016--17 near-IR spectroscopic campaign because of the low-flux state \ngc\ was observed in and the presence of persistent, broad He\,\textsc{i}~$\lambda1.08$~\um absorption associated with the obscurer (\citealt{Wildy21}).

\cite{Mehdipour15} presented two SEDs based on 2013--14 X-ray observations: one seen through the obscurer, and the intrinsic nuclear SED.
We show these SEDs in Fig.~\ref{fig:seds}, where we mark also the ionisation potentials of the near-IR coronal lines.
Clearly the coronal line emission would be strongly affected by the obscurer if it intervenes between the coronal line gas and the X-ray source.
We do not know \textit{a priori} the location of the coronal line emitting gas and therefore we can explore whether the line emission predicted by the obscured or unobscured SED better match our observations. Therefore, we used both SEDs in \cloudy\ to make predictions about the resultant line emission.

\subsection{The \cloudy\ parameter region}\label{sec:cloudy_params}

We used the fluxes of the coronal line cores reported in Table~\ref{tab:mean_fits}\footnote{For \se\ and \silvi\ these are the line core fluxes (excluding the wings).}, and calculated the equivalent widths of the lines relative to the 1215~\si\angstrom\ continuum flux of the \cite{Mehdipour15} SEDs. Subsequently, we searched the parameter space for regions which reproduced the observed emission line equivalent widths within uncertainties of $\approx30$~per~cent.
Because we expect that the \siii-emitting gas is not cospatial with the coronal line gas, the \cloudy\ predictions for \siii\ emission can be used to discriminate between models. For example,
regions that predict substantial \siii\ emission or overpredict the observed flux may be ruled out as potential sites for the coronal line gas. On the other hand, regions that produce little or no \siii\ are still viable, since this emission line flux could come from elsewhere.
We additionally considered the equivalent widths of the optical Fe coronal lines (Section~\ref{sec:optical}) to aid our search.
We note that lines were measured in a spectrum from 2015 March and are therefore non-contemporaneous with our near-infrared data from 2016--17.
Flux variability of a factor $\sim2$ was seen in the [Fe\,\textsc{vii}] lines over $\approx5$~years (\citealt{Landt15b})\textcolor{black}{.}
\textcolor{black}{Given this level of intrinsic variability (and some uncertainty in the absolute flux scaling of the optical spectrum) we reasonably expect the optical coronal lines to have fluxes within a factor $\approx2$--3 of those we measure in the 2015 spectrum.}

Our \cloudy\ models are “luminosity cases” similar to the studies of \cite{Ferg97} and \cite{bald95}. 
This means that we have used the obscured and unobscured SEDs with their corresponding bolometric luminosities estimated by \cite{Mehdipour15}.
In particular, for the unobscured case,
the bolometric luminosity is
$\log(L_{\mathrm{bol}}/\mathrm{erg\,s}^{-1})=44.2$, whereas, for the
obscured case, it decreases to
$\log(L_{\mathrm{bol}}/\mathrm{erg\,s}^{-1})=43.5$. Below, we compare
all results to the observed values of
the coronal lines from the 2016--17
campaign of \cite{Landt19}. While a
three-year time span between the SEDs
and the coronal lines probably slightly
affects the results, it is also possible
that the SEDs were quite different in
2016, which would lead to very different
predictions. For this reason, we choose
not to fine-tune the \cloudy\  results, and
we only propose them as a possible
solution while we emphasize the
importance of the approach. For the
future application of the method, it
would be beneficial to measure both the
emission lines and ionising continuum in the same time period. 

Another source of uncertainty is the
column density: the hydrogen column
density of the possible coronal line
emitting cloud is unknown. The stopping
criteria for the \cloudy\ models is set to
be the column density of the cloud,
which is assumed to be
$\log(N_\mathrm{H}/\mathrm{cm}^{-2})=23$
\textcolor{black}{during most of our
calculations}. This is a typical value
assumed for the coronal line emitting
regions; however, it is not measured and
can be a different value \textcolor{black}{so we have also
created some models using
$\log(N_\mathrm{H}/\mathrm{cm}^{-2})=22$ and 22.5}.
The cloud has an ionisation structure
with the highest ionisation lines
forming near its illuminated face and
the lowest ionisation lines forming near
the shielded face. Decreasing the column
density of the cloud will not  \textcolor{black}{dramatically} affect the
higher ionisation lines but will reduce
the intensity of the low-ionisation
lines. \textcolor{black}{To check this claim, we created two
obscured models with different column
densities ($\log[N_\mathrm{H}/\mathrm{cm}^{-2}]=22$ and 23),  
and similar hydrogen density of  $\log(n_\mathrm{H}/\mathrm{cm}^{-3})=5$. Both
clouds were located at the same distance
of $\log(R/\mathrm{cm})=18$. The \cloudy\  calculations
indeed show that decreasing the column
density by 1 dex (from $\log[N_\mathrm{H}/\mathrm{cm}^{-2}]=23$ to 22) reduces the EW of
high ionization lines only by 15--30~per cent,
while the low ionization lines become at
least 60~per cent weaker.}  
Therefore, we
think that the variation of the column
density does  not \textcolor{black}{strongly} affect the
high-ionisation lines, but does impact on the
low-ionisation lines.
\textcolor{black}{It is worth mentioning
that tests have shown that the cloud can
be highly ionized in some cases. In such
cases, the temperature at the
illuminated face of the cloud gets very high ($>10^{6}$~K in some cases), resulting in almost no
near-IR and UV high-ionization lines being produced at the
illuminated face. 
In those cases, the
EWs of all coronal lines will significantly
change by varying the column density.
}

\begin{figure*}
    \centering
    \begin{tabular}{cc}
        \includegraphics[width=1.15\columnwidth]{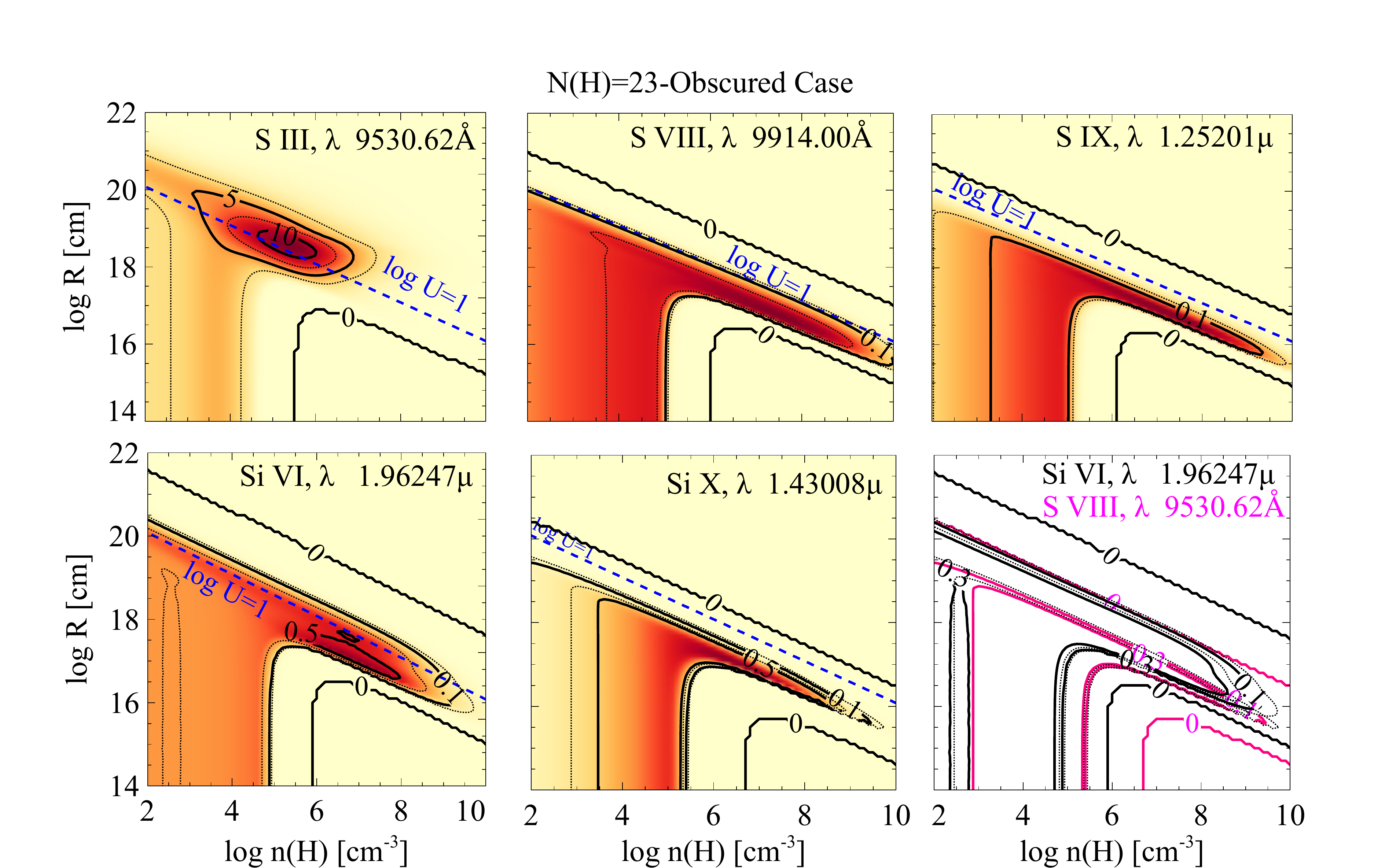} & \includegraphics[width=0.85\columnwidth]{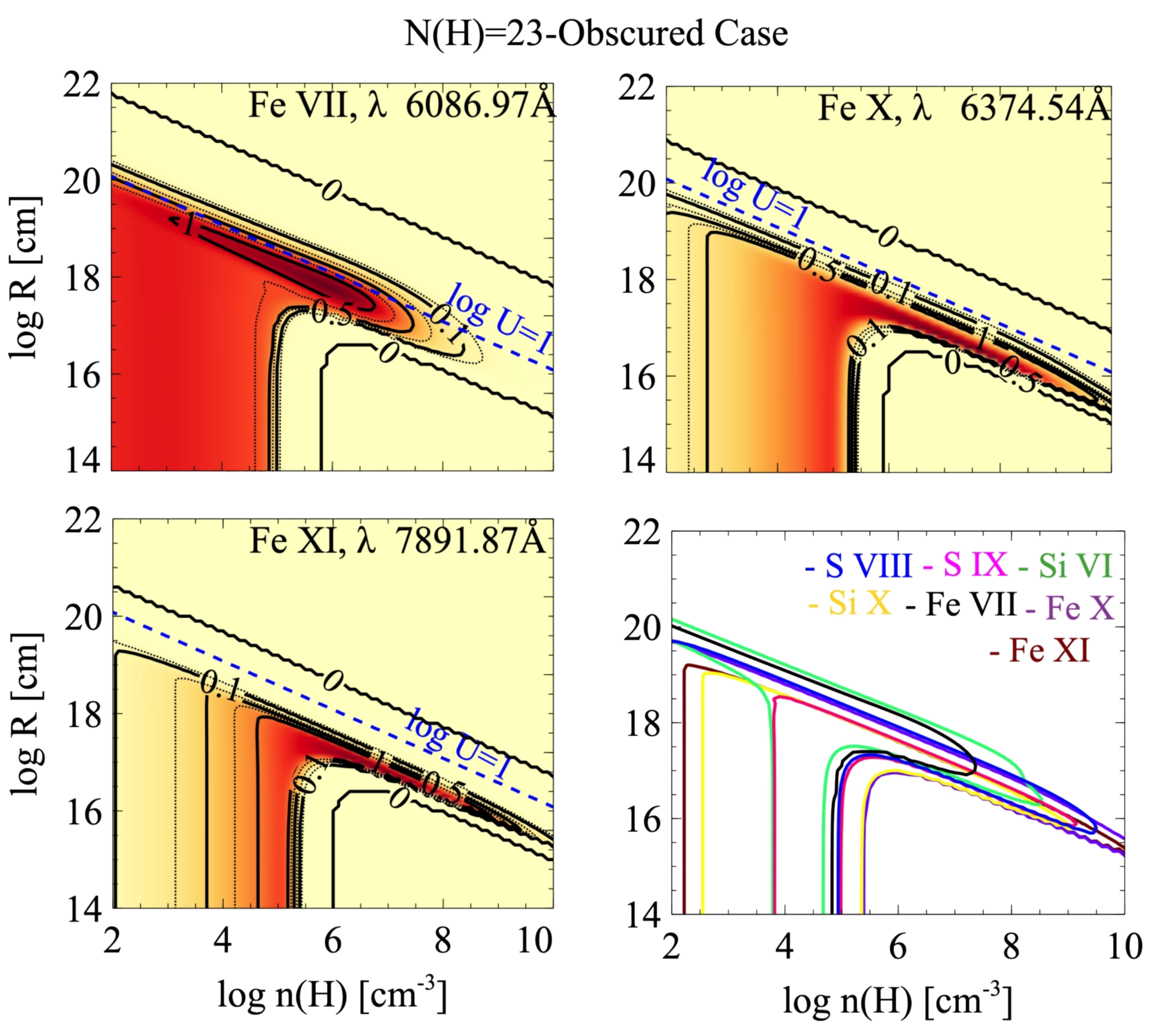} \\
        (a) Radius $R$ versus density $n$ for the obscured SED: S and Si lines. & (b) Radius $R$ versus density $n$ for the obscured SED: Fe lines. \\
        \includegraphics[width=1.15\columnwidth]{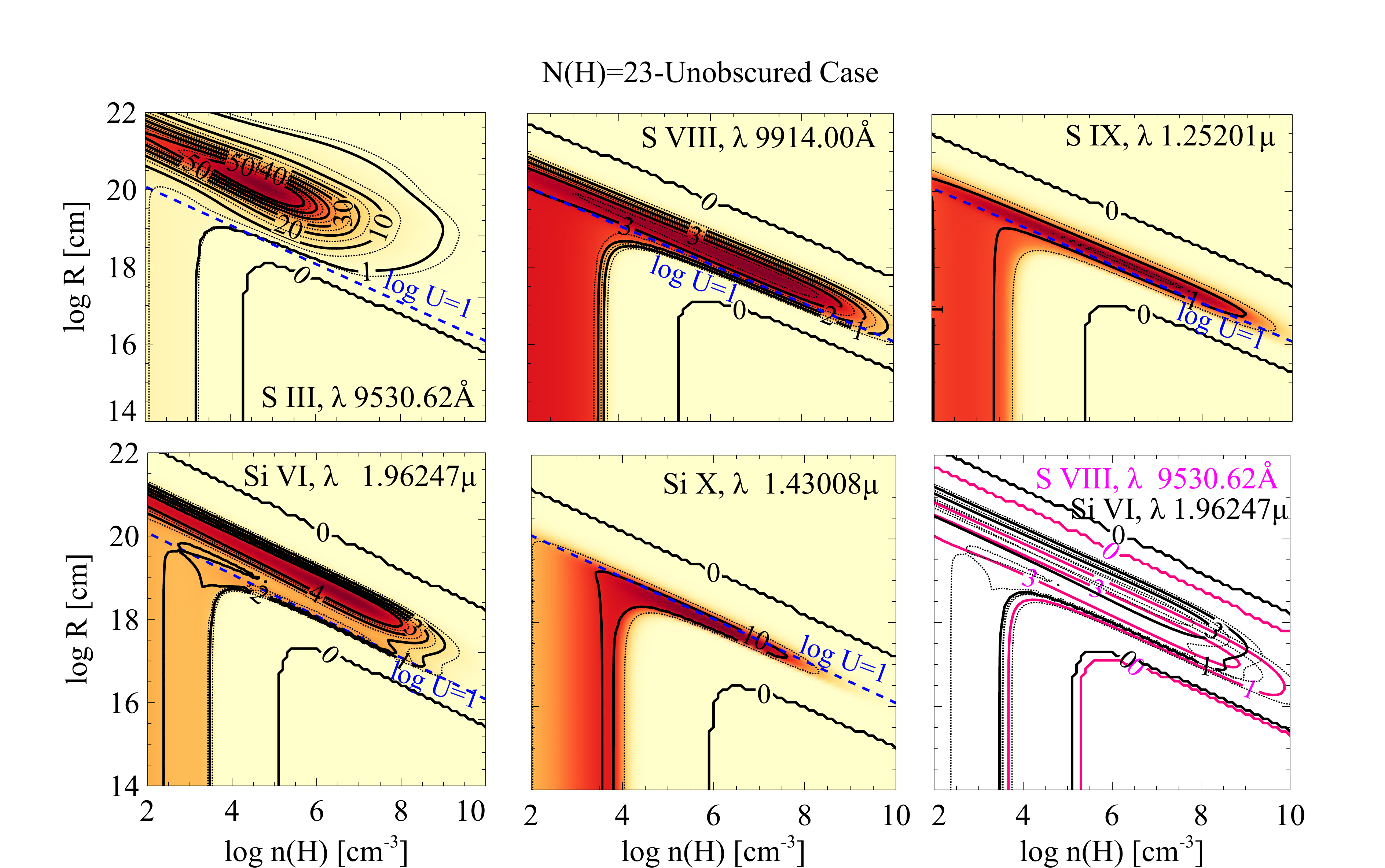} & \includegraphics[width=0.85\columnwidth]{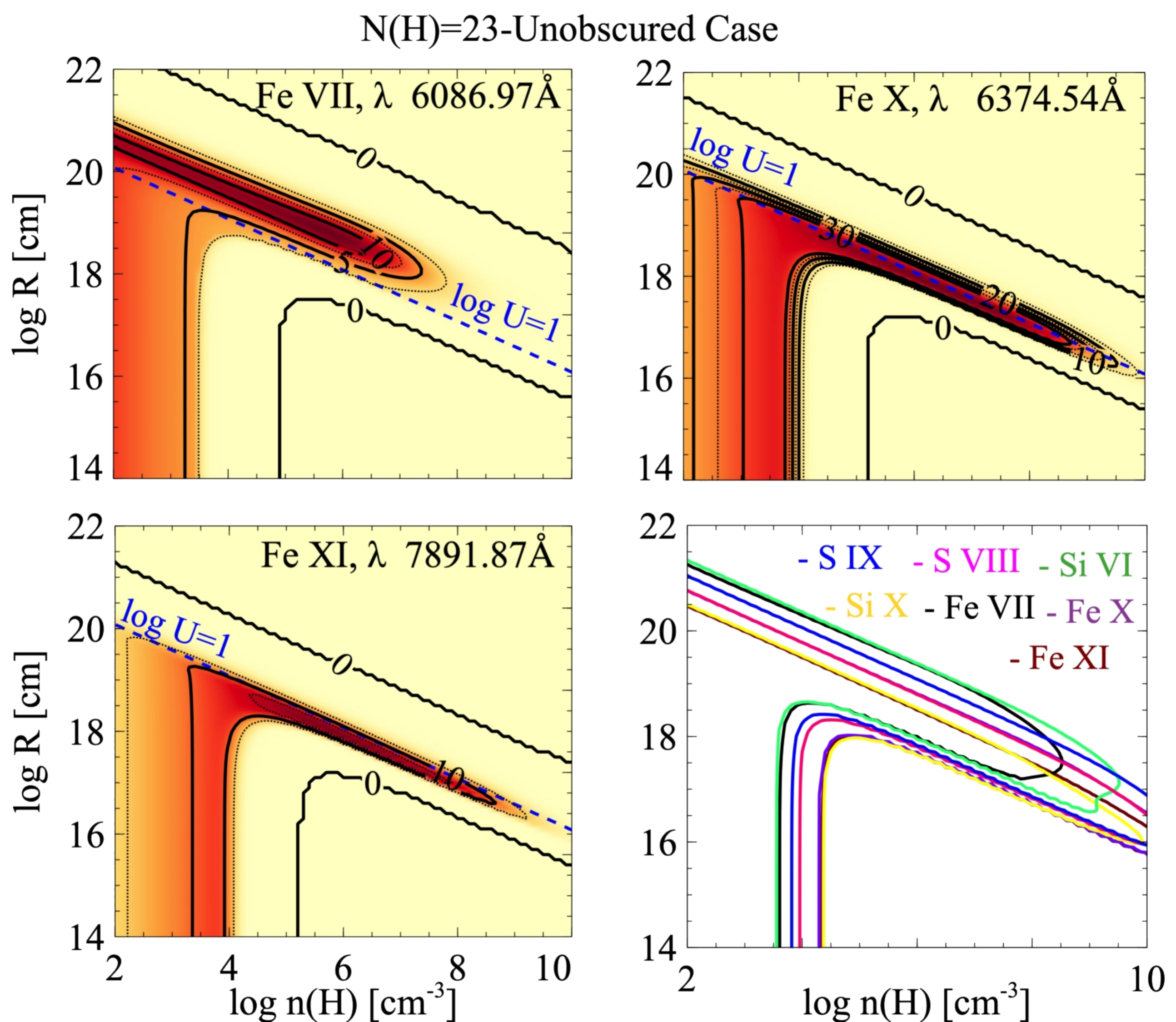} \\
        (c) Radius $R$ versus density $n$ for the unobscured SED: S and Si lines. & (d) Radius $R$ versus density $n$ for the unobscured SED: Fe lines. \\
    \end{tabular}
    \caption{\cloudy\ simulations for the four near-IR coronal lines, three optical coronal lines and \siii~$\lambda9530$. 
    Plots (a) and (c) (plots [b] and [d]) show the radius and density of the S and Si (Fe) line-emitting gas. 
    Contours trace the equivalent widths (EWs) of the emission lines with respect to the continuum at 1215~\AA. 
    The blue dashed line shows the ionisation parameter log $U=1$ in the parameter space. 
    Plots (a) and (b) have been calculated using the obscured SED; plots (c) and (d) using the unobscured SED.
    The bottom right-hand panels in plots (a) and (c) show an overlay of the contours for \se\ and \silvi\ (the two coronal lines which exhibit wings) for easier comparison. 
    The bottom right-hand panels in plots (b) and (d) show\textcolor{black}{s} the EW contours corresponding to the observed fluxes of the optical coronal lines in a single panel.
    }
    \label{fig:cloudy}
\end{figure*}

Fig.~\ref{fig:cloudy} shows the results for \textcolor{black}{the models with
$\log(N_\mathrm{H}/\mathrm{cm}^{-2})=23$}. The
obscured case is shown in plots \ref{fig:cloudy}(a) and (b), and the unobscured case is shown in plots (c) and (d). 
Plots (a) and (c) illustrate how the equivalent width of each near-IR coronal line (and also \siii) depends on the location (vertical axis) and the hydrogen density (horizontal axis) of the cloud. 
The lower-right panel in these plots show the contours for \se\ and \silvi\ overlaid. 
Plots (b) and (d) show the contours for the optical coronal lines ({[Fe\,\textsc{vii}]} 3759\AA, {[Fe\,\textsc{x}]} 6374\AA, and {[Fe\,\textsc{xi}]}  7892\AA). The lower-right panels of these series of plots have been created using the approach introduced by \cite{Dehghanian20}. 
In these panels, each coloured line shows the observed value of one specific coronal line. By tracing these lines, one might be able find the place where they (almost) cross over each other. The crossover point would indicate the location and the density of the cloud that produces the observed line strengths.

\subsection{Solution for the coronal line cores}
\label{sec:cloudy-cores}
Considering Fig.~\ref{fig:cloudy}, we ran over a hundred
location-density points for three
column densities of
$\log(N_\mathrm{H}/\mathrm{cm}^{-2})=22$, $22.5$ and $23$.
Tables~\ref{tab:cloudyObs} and
\ref{tab:cloudyUno} show
\textcolor{black}{some of the results from
the mentioned \cloudy\ simulations. We
mainly selected those clouds
that produce zero to one
hundred per~cent of near-IR coronal lines within
a reasonable range of
uncertainty (a tolerance of $\approx30$~per cent). Models 3 and 5
overproduce \se\  beyond this uncertainty,
but are still presented in the table since
they produce the right amount of \silvi.} 

\textcolor{black}{Regarding the optical
coronal lines (\feii, \fevii\ and \fex), as explained in Section~\ref{sec:cloudy_params},
since we lack contemporaneous measures of these lines
we search for models which agree with the observed values to within a factor $\approx2$--3.}

As both tables indicate, there is not a single location-density solution for which the cloud produces all of the observed coronal lines, implying that we must be seeing emission from different regions. By comparing the models, we note that while \se\ and \silvi\ are produced by the same cloud, \sn\ and \sit\ are emitted by different cloud(s). Also, since we could find more solutions in the obscured case, it is most likely that the coronal regions saw the photonionising source through the obscurer during this near-IR spectroscopic campaign. 
However, it is also possible that the continuum source was unobscured to some coronal line production sites (i.e.\ the coronal line gas was located between the obscurer and the source), but obscured at other locations (i.e.\ the gas was located between the obscurer and the observer).
The \cloudy\ simulations indicate that models
3 (or 5) and 7 are most likely responsible
for producing the observed values of the
near-IR coronal line cores. In this scenario,
cloud 7 produces almost all of the \sit,
while cloud 3 (or 5) produces \se, \sn\ and
\silvi. \textcolor{black}{Model 4 is also a good choice to produce the right amount of \se\ and \silvi, however this cloud does not produce enough \sn.}

In general, it is difficult to
produce sufficient \sit~emission without
underestimating the emission from the other
near-IR coronal lines. Furthermore, the
\sit~region requires on average a gas density
reduced by a factor of $\sim 1000$ (and a
somewhat larger distance from the central continuum source).

\begin{table*}
    \centering
    \caption{A comparison of \cloudy\ coronal line flux predictions to the observed values in the obscured case}
    \label{tab:cloudyObs}
    \begin{tabular}{llccccccc}
    \hline
    \multicolumn{2}{c}{Coronal line}                  & model 1 & model 2 & model 3 & model 4 & model 5         & model 6 & model 7  \\
                        & & $N_\mathrm{H}=10^{23}$     & $N_\mathrm{H}=10^{22}$   & $N_\mathrm{H}=10^{22.5}$              & $N_\mathrm{H}=10^{22.5}$  & $N_\mathrm{H}=10^{22.5}$ & $N_\mathrm{H}=10^{23}$ & $N_\mathrm{H}=10^{22}$
                         \\
                         & &$R=10^{17.3}$
                          &$R=10^{16.3}$ 
                          &$R=10^{16.6}$ 
                          &$R=10^{16.7}$ 
                          &$R=10^{16.6}$ 
                          &$R=10^{17.6}$ 
                          &$R=10^{18}$
                          \\
                        &  &$n_\mathrm{H}=10^{7.4}$
                        &$n_\mathrm{H}=10^{8}$
                        &$n_\mathrm{H}=10^{8.3}$
                        &$n_\mathrm{H}=10^{8.3}$ 
                        &$n_\mathrm{H}=10^{8.2}$
                        &$n_\mathrm{H}=10^{7.5}$
                        &$n_\mathrm{H}=10^{4.5}$
                        \\
    \multicolumn{2}{c}{(1)}                 & (2)       & (3)    & (4)                     & (5)             & (6)           & (7) & (8)\\
    \hline
    {\se} & 0.9914~\um & 70\% & $<1$\% & 151\% & 118\% & 167\% & 3\% & $<1$\% \\
    {\sn} & 1.2520~\um & 7\% & 2\% & 46\% & 23\% & 64\% & $<1$\% & 4\% \\
    {\silvi} & 1.9625~\um & 129\% & 0\% & 81\% & 79\% &  90\% &  58\% & $<1$\% \\
    {\sit} & 1.4300~\um & $<1$\% & 39\% & 5\% & 2\% &  12\% &  0\% & 81\% \\
    {[Fe\,\textsc{vii}]} & 3759\AA & 78\% & $<1$\% & 19\% & 18\% &  23\% &  20\% & $<1$\% \\
    {[Fe\,\textsc{x}]} & 6374\AA & 23\% & 97\% & 188\% & 77\% &  309\% &  $<1$\% & 212\% \\
    {[Fe\,\textsc{xi}]}  & 7892\AA     & $<1$\% & 81\% & 10\% & 3\% &  20\% &  $<1$\% & 167\% \\
    \hline
    \end{tabular}
\parbox[]{14.8cm}{The columns are: (1) Ion and wavelength of the near-infrared transition; (2)--(8) the percentage of the observed flux predicted by different \cloudy\ models.
\textcolor{black}{For \se\ and \silvi\ the percentages are relative to the line core fluxes (excluding the wings).}}
\end{table*}
\begin{table}
    \centering
    \caption{A comparison of \cloudy\ coronal line flux predictions to the observed values in the unobscured case}
    \label{tab:cloudyUno}
    \begin{tabular}{llccc}
    \hline
    \multicolumn{2}{c}{Coronal line}                  & model 8 & model 9 & model 10 \\
                         & & $N_\mathrm{H}=10^{23}$     & $N_\mathrm{H}=10^{23}$   & $N_\mathrm{H}=10^{23}$ 
                         \\
                         & &$R=10^{17.25}$
                          &$R=10^{16.2}$ 
                          &$R=10^{17.55}$ 
                          \\
                         & &$n_\mathrm{H}=10^{9.5}$
                        &$n_\mathrm{H}=10^{4.4}$
                        &$n_\mathrm{H}=10^{6.3}$
                        \\
    \multicolumn{2}{c}{(1)}                & (2)       & (3)    & (4)\\
    \hline
    {\se}                & 0.9914~\um & 92\%   & 0\%     & $<1$\% \\
    {\sn}                & 1.2520~\um & 3\%    & $<1$\%  & $<1$\%\\
    {\silvi}             & 1.9625~\um & 94\%   & 0\%     & 0\% \\
    {\sit}               & 1.4300~\um & 0\%    & 58\%    & 110\% \\
    {[Fe\,\textsc{vii}]} & 3759\AA             & 8\%    & 0\%     & 0\% \\
    {[Fe\,\textsc{x}]}   & 6374\AA             & 3\%    & 5\%     & 10\% \\
    {[Fe\,\textsc{xi}]}  & 7892\AA             & $<1$\% & 13\%    & 35\% \\
    \hline
    \end{tabular}
\parbox[]{8cm}{The columns are: (1) Ion and wavelength of the near-infrared transition; (2)--(4) the percentage of the observed flux predicted by different \cloudy\ models.
\textcolor{black}{For \se\ and \silvi\ the percentages are relative to the line core fluxes (excluding the wings).}}
\end{table}

\subsection{Solution for the coronal line wings}
\label{sec:cloudy-wings}
\textcolor{black}{The values listed in Tables \ref{tab:cloudyObs} and \ref{tab:cloudyUno} are fluxes relative to the mean flux of the narrow emission line cores.
In our mean spectrum, the wings of \se\ and \silvi\ are $62\pm11$ and $95\pm7$~per cent of the core fluxes, respectively.
The ideal model to explain the coronal line wings will therefore produce a flux in \se\ equivalent to $\approx62$~per cent of its core flux, a flux in \silvi\ equivalent to $\approx95$~per cent of its core flux, and 0~per cent of the core flux of all of the other lines (none of which display wings).
}
For the unobscured SED, we find a solution (model 8, Table~\ref{tab:cloudyUno}) at a radius $\log(R/\mathrm{cm})=17.25$ which produces fluxes in \silvi\ and \se\ that are very similar to the observed fluxes in the broad wings of these lines.
The model predicts a flux equivalent to 94~per~cent of the narrow core flux of \silvi\ will be produced at this location (precisely what we measure for the flux in the wings in the mean spectrum). It also predicts an equivalent of 92~per~cent of the \se\ flux will be produced here (an overprediction, since we measure 62~per~cent). 
Furthermore, no \sit\ and very little \sn\ emission (\textcolor{black}{equivalent to} 3~per~cent of the core flux) are produced in this region.
\textcolor{black}{Model 8 is therefore a near-ideal solution for the coronal line wings.}
Model 1 in the obscured case (Table~\ref{tab:cloudyObs}) has some similar properties to model 8: it is at a similar radius ($\log[R/\mathrm{cm}]=17.3$), and produces substantial \se\ and \silvi\ emission, with little \sn\ and \sit. 
\textcolor{black}{However,} more Fe coronal line emission is produced in this model compared with model 8: $\approx78$~per~cent of the observed [Fe\,\textsc{vii}]$\lambda3759$ and $\approx23$~per~cent of the observed [Fe\,\textsc{x}].
Interestingly, the radius of both of these coronal line emitting clouds $\log(R/\mathrm{cm})\approx17.3$ or 70~light~days is exactly the radius of the inner torus as determined from the hot dust reverberation lags by \cite{Landt19}. However, the coronal line gas density in the unobscured case is similar to what is expected for the dusty torus material, whereas it is a factor of $\sim 100$ lower in the obscured case. We discuss this case further in Section~\ref{sec:discussion-wings}.

\section{Discussion}
\label{sec:discussion}
We have performed an in-depth study of the near-IR coronal lines in \ngc, making use of spectroscopic data recorded over a year with a roughly weekly cadence.
We were able to study not only the line profile shapes but the changes in line shapes and fluxes.
We have shown that two of the coronal lines (\se\ and \silvi) have prominent broad wings as well as narrow cores whereas only narrow cores are evident on the other two coronal lines.
Whilst the narrow line cores are persistent and clearly detected in all spectra, the wings are highly variable; the flux in the wings is comparable to the flux in the line cores in some epochs and in other epochs the wings are barely visible.
In the case of \silvi\ it is clear that most of the variability in the line profile is in the wings.
Broadly speaking, there must be at least two coronal line regions in \ngc: one which produces the persistent, narrow cores of all four coronal lines and another (presumably more compact) region in which the conditions favour the production of \se\ and \silvi\ emission which we observe as the broad, variable wings.

\begin{figure}
    \centering
    \includegraphics[width=\columnwidth]{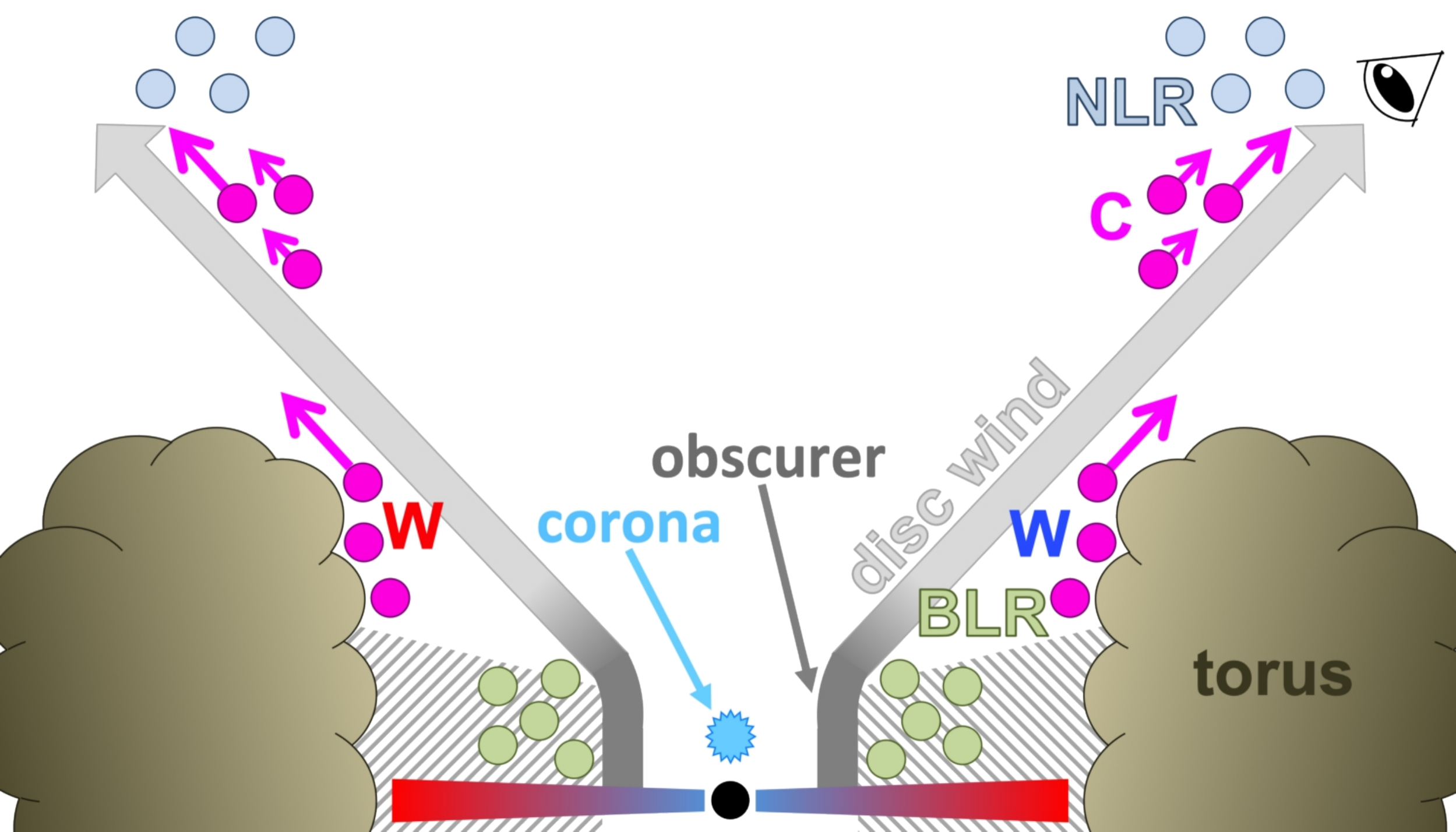}
    \caption{A cartoon of the possible geometry of the nucleus in \ngc.
    A wind \protect{\citep{Dehghanian19}} is launched from the disc: its dense base (the `obscurer' in dark grey) partly shadows the BLR (light green clouds).  The less dense streamlines of the wind (light grey) extend to large distances.
    Gas producing the narrow cores of the coronal lines (magenta clouds labelled `C') is located just inside of the standard NLR (light blue clouds).
    The outflowing coronal line gas may be accelerating, as indicated by the magenta arrows.
    The blue and red wings seen on the \se\ and \silvi\ coronal lines (marked with the blue and red `W', respectively) may be produced on the inner face of the torus.
    The torus may partially block our view of the gas producing the blue wing.}
    \label{fig:cartoon}
\end{figure}

The existence of multiple sites of coronal line emission in AGN has been proposed before based on comparisons between Seyfert type 1 and type 2, but never before shown for a single object. \cite{Mur98} reported excess [Fe\,\textsc{vii}]~$\lambda6087$ emission in Seyfert 1 nuclei compared with Seyfert 2s, implying that some coronal line emission occurs interior to the dust torus (and so can be seen only in Seyfert 1s) and some coronal line emission occurs beyond the torus.
They proposed three main coronal line regions: the inner face of the dust torus, highly-ionised clumps of gas in the NLR and a further, very extended region on kpc-scales. We take up this interpretation for the coronal line regions in \ngc\ in more detail below and sketch it in Fig. \ref{fig:cartoon}.

\subsection{The coronal line cores}
\label{sec:discussion-cores}
\begin{figure}
    \centering
    \includegraphics[width=\columnwidth]{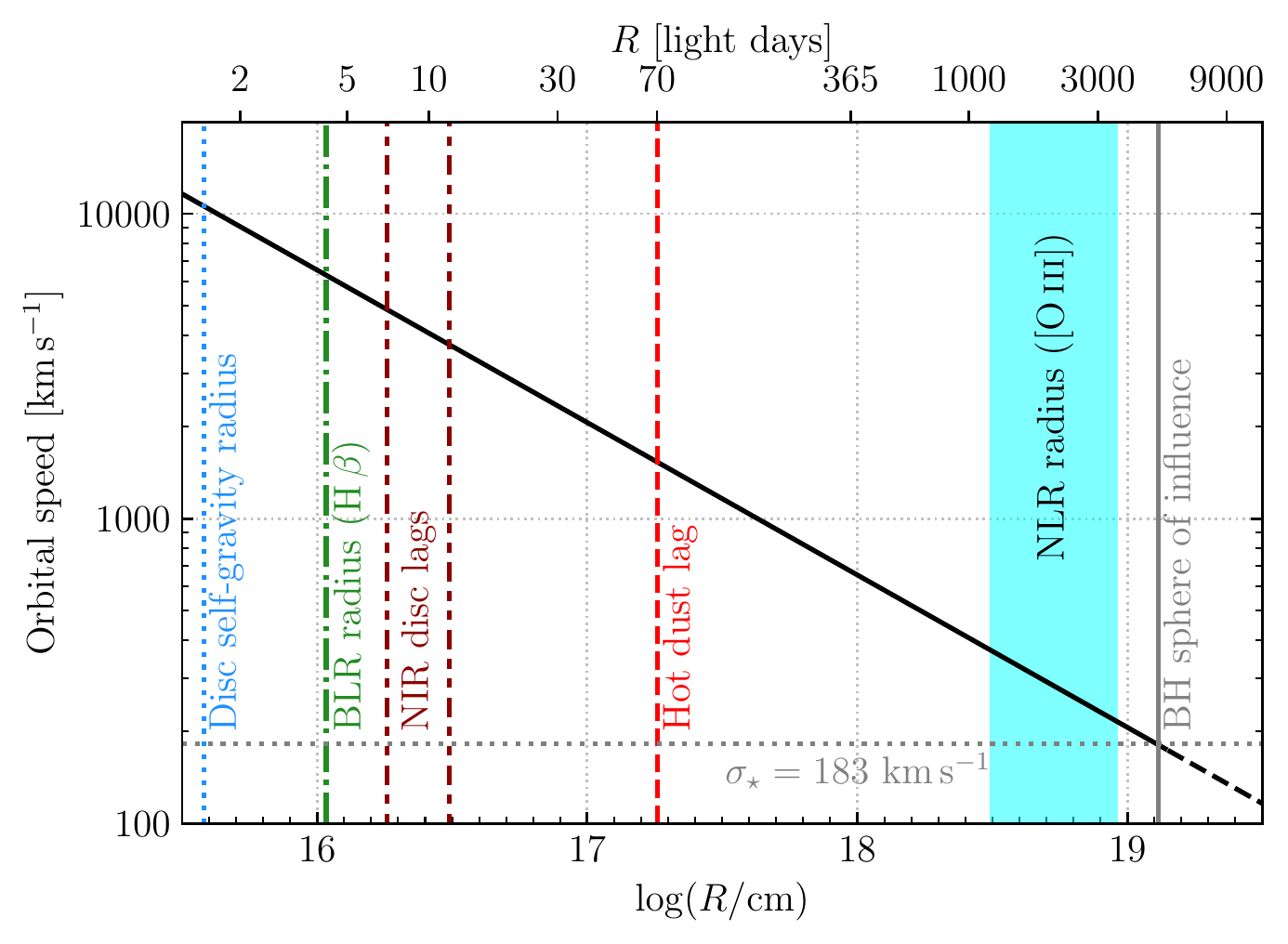}
    \caption{Orbital velocity as a function of radius for a black hole of mass $3.2\times10^{7}$~M$_\odot$.
    The black hole's sphere of influence extends up to $\log(R/\mathrm{cm})\approx19.1$.
    We indicate radii of the H\,$\upbeta$ BLR (green: \citealt{Pei17}); hot dust (red: \citealt{Landt19}) and \oiii\ NLR (cyan: \citealt{Pet13}).
    $H$ and $K$ band lags attributed to the outer accretion disc (dark red: \citealt{Landt19}) suggest that the disc extends beyond its self-gravity radius (blue: \citealt{Gardner17}).
    The X-ray obscurer (the base of a disc wind) must be interior to the BLR \citep{Dehghanian19}. 
    The stellar velocity dispersion of the bulge was measured to be $\sigma_\star=183$~\kms\ \protect\citep{Ferrarese01}.}
    \label{fig:rad_vel}
\end{figure}

In Section~\ref{sec:meanspec} we carefully analysed the emission line profiles in the high S/N mean spectrum.
In addition to the four near-infrared coronal lines we also measured the low-ionisation forbidden line \siii$\lambda9531$.
In comparison to \siii, the cores of all four coronal lines are broader (ranging from $\mathrm{FWHM}\approx550$--750~\kms) and are blueshifted with respect to it by $\approx120$--380~\kms\ (Fig.~\ref{fig:coronal_line_Gauss}).
The greater widths and blueshifts of the coronal lines with respect to the low-ionisation forbidden lines is commonly seen in AGN spectra and was reported in early studies (e.g.\ \citealt{Grandi78}; \citealt{Pelat81}).
This implies that the coronal lines are produced in different gas to the standard NLR responsible for the emission from low-ionisation species such as \siii\ and \oiii.
This is consistent with the idea that the coronal line emitting gas is part of an outflowing wind on scales more compact than the standard NLR.
\textcolor{black}{Our measurements of the coronal line cores over the duration of the campaign indicate that they only weakly variable, if at all.
In Section~\ref{sec:variability-trends} we noted a positive relationship between the FWHM and flux over time in the single-epoch measurements of the line cores.
This trend (if real) could be explained if the gas producing the broadest part of the line profiles is more compact (and varies more rapidly) than the gas producing the more persistent narrow cores of the lines. 
However, in our data the same FWHM-flux trend can alternatively be explained by the movement of the underlying continuum within its uncertainty range, so we cannot draw any strong conclusion on this point.
In any case, the weakness of the variability is consistent with a geometry in which majority of the flux in the coronal line cores is emitted at radii $\gtrsim1$~light year and therefore the line cores have a variability timescale longer than we can probe with this data.
Then, considering both the profile widths and variability properties, we may deduce that the coronal line cores are produced at a characteristic radius $18\lesssim \log(R/\mathrm{cm}) \lesssim18.5$ from the nucleus, a scale intermediate between the BLR and NLR (Fig.~\ref{fig:rad_vel}).}

Although the widths and shifts of the four coronal line cores have common characteristics (being broader and blueshifted with respect to \siii), they are not identical.
It is likely that the lines are not emitted entirely cospatially and different coronal lines originate in different parts of the outflow.
Our photoionisation simulations with \cloudy\ (Section~\ref{sec:cloudy}) also strongly suggest this, since we do not find any single cloud that can account for all of the observed lines.
In the mean spectrum, \sn\ was found to be the narrowest coronal line ($\mathrm{FWHM}=554\pm24$~\kms) and \se\ the broadest ($\mathrm{FWHM}=737\pm39$~\kms).
Consistent with previous studies (e.g. \citealt{Rod11}; \citealt{Ferg97}; \citealt{DeRobertis86}; \citealt{Filippenko84} \citealt{Pen84}; \citealt{Pelat81}), we found a trend of increasing FWHM with the ionisation potential of forbidden emission lines (Fig.~\ref{fig:ip})\footnote{Contrary to \cite{Filippenko84} we observe a weaker relationship between emission line FWHM and critical density, as can be seen in Fig.~\ref{fig:ip}.}.
\cite{Rod11} noted that this trend extended up to $\chi\approx~300$~eV, above which the FWHM of lines decreased or plateaued, which is what we observe in Fig.~\ref{fig:ip}: \sit\ ($\chi=351.1$~eV) is no broader than \se\ ($\chi=281.0$~eV) and \sn\ ($\chi=328.8$~eV) is narrower than \se.
Since \se\ has the highest critical density of the lines studied here, some of its emission may arise from higher-density gas nearer to the nucleus where orbital motions are greater.  
In this picture it is therefore expected that \se\ will be the broadest coronal line, as we observe.
This result suggests a degree of stratification of the coronal line emitting gas.

\cite{Pelat81} found a relationship between line width and ionisation potential, as well as line width and velocity shift, with the general trend that higher-ionisation lines were broader and more strongly blueshifted.
We see such a trend to first order (the coronal lines are broader than, and blueshifted with respect to, \siii); however, the reverse trend is seen within the coronal line sequence: the lowest-ionisation coronal line \silvi\ has the greatest blueshift (Fig.~\ref{fig:ip}).
The \se, \sn\ and \sit\ lines have centroid shifts that are consistent within error, with a mean and standard deviation of $-125$ and 9~\kms, respectively; the \silvi\ line has a significantly greater shift of $-383\pm10$~\kms.
This difference in blueshifts is real, and not the result of a systematic error in the wavelength calibration at the red end of our spectra: we assessed the centroid shifts of the low-ionisation \feii$\lambda1.644$~\um line near to \silvi\ and did not find any significant shift of this line.
Therefore the kinematics of the \silvi-emitting gas appear to be different to that of gas emitting the other three coronal lines.
\cite{Ferg97} noted that their photoionisation models predicted that lower-ionisation coronal lines ([Ne\,\textsc{v}]--[Si\,\textsc{vii}]) would form in more extended gas than the high-ionisation lines.
We see this trend in our \cloudy\ simulations, in which the contours of \silvi\ are displaced to larger radii compared with the other coronal lines (see the comparison of \se\ and \silvi\ contours in Fig.~\ref{fig:cloudy}).
We also find that the core of the \silvi\ line is narrower than that of \se\ and \sit, which (if interpreting the line widths as orbital motion) would also imply its emission occurs at typically larger radii.
If the \silvi-emitting gas is located at larger radii and greater velocities than the rest of the coronal line gas, this implies that the gas is part of an accelerating outflow.
Outflows which accelerate from the nucleus to distances of $\sim100$~pc have been observed in several other AGN (e.g.\ \citealt{Crenshaw00a}; \citealt{Crenshaw00b}; \citealt{Mueller11}).

Our photoionisation simulations appear to predict a very compact origin for the coronal line cores (Section~\ref{sec:cloudy-cores}).
\cloudy\ models 3 or 5 produce approximately the same fluxes as we observe in the cores of the \se, \sit\ and \silvi\ lines and the location of the line-emitting gas in these solutions is $\log(R/\mathrm{cm})=16.6$, a similar scale to the BLR or outer accretion disc (see Fig.~\ref{fig:rad_vel}).
The widths of the lines do not necessarily reflect Doppler broadening by virial motion of the emitting gas; instead the lines may be broadened by gas turbulence as is the case in coronal line novae.
However, the blueshifts of the coronal lines strongly suggest that the emitting gas is outflowing.  
It is then reasonable to assume that the outflow is launched from the rotating accretion disc. 
In this case we would expect the gas to have a large amount of rotational as well as radial motion, and for this to be evident in the widths of the emission lines.

The \cloudy\ simulations also suggest that the three coronal lines \se, \sn\ and \silvi\ may be produced in the same cloud (either 3 or 5), and \sit\ is produced in another cloud (7). 
However, the most obvious difference in the coronal line core profiles is the much greater blueshift of \silvi\ with respect to the other lines. 
This suggests that \silvi\ emitting gas is kinematically (and spatially) distinct from the gas producing the other three coronal lines, or alternatively has an additional component not seen in the other lines, which would not be accounted for in our \cloudy\ solutions. 

\cite{Landt15b} determined greater distances for the coronal line gas in \ngc\ than implied here.
Based on measurements of the optical and X-ray coronal lines, they calculated that the coronal lines were produced in a low-density gas ($\log[n_\mathrm{e}/\mathrm{cm}^{-3}]\sim3$) at $\log(R/\mathrm{cm})\approx18.9$ ($\approx8$~light~years).
This distance is approximately coincident with the \oiii~$\lambda5007$ NLR and is just within the black hole's gravitational sphere of influence (see Fig.~\ref{fig:rad_vel}). 
In spite of the similar size scale to the NLR, \cite{Landt15b} proposed that the coronal line region was likely to be an independent entity because of its high ionisation parameter ($U\sim1$).
The modest variability of the coronal line cores and their relatively narrow widths imply their origin in gas interior to the standard NLR ($\log[R/\mathrm{cm}]\lesssim18$).
The photoionisation simulations in Section~\ref{sec:cloudy-cores} suggest an even more compact coronal line region at $\log(R/\mathrm{cm})=16.6$.
It can be seen from Tables~\ref{tab:cloudyObs} and \ref{tab:cloudyUno} that \cloudy\ generally predicts much higher gas densities ($\log[n_\mathrm{H}/\mathrm{cm}^{-3}]\sim7$--9) than calculated by \cite{Landt15b}.
Our results are not necessarily at odds with those of \cite{Landt15b} since they derived physical parameters from optical and X-ray coronal lines, whereas we do so for the near-IR coronal lines. Since emission lines form wherever the conditions allow them to do so, it is likely that different coronal lines trace clumps of differing density, thus mapping out the outflowing wind that produces them. 

In summary, the kinematics inferred from \textcolor{black}{the narrow line profile widths} suggest that the coronal line cores are emitted in an accelerating wind just interior to the low-ionisation NLR. However, at odds with this interpretation, we did not find photoionisation model solutions consistent with it, although there are a number of limitations to this modelling, as mentioned earlier. Therefore, it is possible that the line widths are not produced by virial motion within the black hole's gravitational field, but rather indicate radial motion as in the ejecta of classical novae during their coronal line phase \citep[e.g.][]{Greenhouse90, Woodward21}.

\subsection{The coronal line wings}
\label{sec:discussion-wings}
\textcolor{black}{Before proceeding, is worth considering whether the flux excesses seen near \silvi\ and \se\ (their `wings'), and the apparent variability of these features, could alternatively be explained by other spectral components unrelated to the coronal lines (i.e.\ the continuum or the blended broad emission lines).
We have incorporated the uncertainty on the placement of the continuum (assumed to be locally linear) in our flux measurements and the variability of the excess flux redward of \silvi\ is still highly significant; therefore, we can rule out the excess flux changes being due to the imperfect subtraction of a (smooth) continuum. 
Of course, the shapes of these flux excesses do not look like parts of a smooth continuum and it does not seem likely that the variations could be due to variations in unexplained, local features in the continuum, either.
As the spectral decomposition by \cite{Landt19} showed, the continuum beneath \se~$\lambda0.9914$~\um\ is dominated by the accretion disc, whereas the continuum beneath \silvi~$\lambda1.9650$~\um\ is dominated by emission from the hot dust.
So even if it were the case that the wings are in fact continuum features, this interpretation requires there to be some physical link between the disc and dust emission at the \textit{specific} wavelengths $0.9914$ and $1.9650$~\um, respectively, which creates the similarly-shaped and similarly-varying flux excesses.
So we do not think it likely that the appearance and variability of these features can be attributed to the dust or disc continua.}

\textcolor{black}{The other possibility is that these spectral and temporal features are caused by the imperfect deblending of the coronal lines from the broad emission lines.
We used the mean \pab\ profile as a (scaled) template for all of the broad hydrogen lines in our extraction of the coronal line profiles from the single-epoch spectra (Section~\ref{singlespec}). 
Therefore, any deviations of the broad lines from the average profile shape can cause features in the residuals which we may attribute to the coronal lines.  
In the case of \silvi, such an effect is likely to be relatively minor.
Our spectral decomposition of the high-S/N mean spectrum around that line (Fig.~\ref{fig:mean_spec_fits}) indicates that \brd\ and particularly \paa\ are very weak features beneath \silvi. 
The flux in \brd\ beneath the \silvi\ red wing is $\approx6\times10^{-15}$~erg\,s$^{-1}$\,cm$^{-2}$, less than the average flux in the red wing itself ($8.7\times10^{-15}$~erg\,s$^{-1}$\,cm$^{-2}$).
The flux variations we attribute to the \silvi\ red wing are of the order $10\times10^{-15}$~erg\,s$^{-1}$\,cm$^{-2}$.
So, if this variability were actually dominated by variations in \brd\ it would require incredibly large changes in the red wing of \brd.
We would then expect to see concomitant changes in the other broad hydrogen line profiles, which we do not.
So it appears that both \brd\ and \paa\ are too weak in this part of the spectrum to satisfactorily explain the changes we see in the flux excess.
Relatively minor changes in the broad \pad\ profile could more plausibly induce a feature that we mistake for \se\ emission.
But, in this interpretation it is then a coincidence that this flux excess on the blue shoulder of \pad\ appears at just the right location to give the appearance of a red wing on \se, with a similar shape and scale to the excess around \silvi\ (which sits on the red wing of \brd).
Additionally, if our deblending method did not generally work well, it is curious that we do not observe any excess flux around \sn\ (which is blended with broad \pab).
This similarity in the profile shapes and lightcurves of the \se\ and \silvi\ wings actually strengthens our preferred interpretation of these flux excesses as coronal line emission, since our \cloudy\ simulations demonstrate that emission from both \se\ and \silvi\ can be produced by the same gas cloud, on $\sim$light month scales from the nucleus.
We therefore conclude that the observed excess emission around the \se\ and \silvi\ lines (and the variability of these features) is genuinely associated with those lines.}

It is intriguing \textcolor{black}{then} that we see broad wings on two of the near-IR coronal lines (\se\ and \silvi) but not on the other two (\sn\ and \sit).
Since we observe wings on one sulphur and one silicon line, the absence of wings on some lines cannot simply be related to the absence of that element at that location (for example, if silicon were depleted onto dust). Our \cloudy\ models show that the appearance of these wings may be explained by photoionisation: we observe wings on the two coronal lines with the lowest ionisation potentials simply due to the combination of bolometric luminosity of the central source and the gas density of the coronal line emitting gas. 
Models 8 and 1, which we described in Section~\ref{sec:cloudy-wings}, are appealing because they explain a number of observed properties.  
\cloudy\ predicts emission from \silvi\ and \se\ emission at similar fluxes to those we observe in the wings occurring at $\log(R/\mathrm{cm})\approx17.3$. Additionally, very little \sn\ and \sit\ emission is produced at this location, explaining the absence of wings on these two lines.
In model 8 none of the other lines we have assessed (\siii\ or the optical Fe coronal lines) are strongly emitted, either.
Since we observe strong variability in the wings as they disappear and reappear over the year-long campaign, we can infer that the emission must arise on $\sim$light~month scales; the radius $\log(R/\mathrm{cm})\approx17.3$ equates to $\approx70$~light days so the implied spatial scale is consistent with the observed variability timescale. As can be seen in Fig.~\ref{fig:rad_vel}, orbital motions of a few thousand \kms\ are expected at this radius which would explain the observed extents of the wings in velocity. A value of 70~light~days is also the precise radius of the inner edge of the torus determined by \cite{Landt19} via the reverberation of the dust in response to the optical accretion disc emission. The strong similarity in the shapes of the \silvi\ wing and hot dust light-curves (Fig.~\ref{fig:dust_wings}) means that the two will have similar reverberation lags and so it is likely that the hot dust and coronal line wings are emitted at the same radius from the nucleus.

It is clear from Figs.~\ref{fig:coronal_line_Gauss} and \ref{fig:min_max_wings} that the broad coronal line wings are much more prominent on the red side of the core than the blue side.
If the wings are indeed produced in a wind launched off the inner edge of the torus, then the torus may be inclined such that our view of the approaching gas is blocked, and we see mostly the receding gas on its far side (see Fig. \ref{fig:cartoon}). A similar geometry was proposed also by \cite{Glidden16} to explain observations of coronal line forest AGN (see their figure 1).
\cite{Pier95} proposed that AGN coronal line emission would originate in an X-ray heated wind evaporated from the inner edge of the dust torus, and this was one of the main sites of coronal line production outlined by \cite{Mur98}. Our findings are consistent with these schemes. However, our chosen model is problematic in that the torus must see the unobscured SED, yet the obscurer is located between the torus and the X-ray source (see e.g.\ figure 4 of \citealt{Kaastra14}). A possible solution is that, if the obscurer is an extended stream or wind, it presents a much higher column density along our line of sight than toward the torus. In any case, the torus emission is not sensitive to the obscurer since it is heated mainly by the UV/optical nuclear emission and the obscurer is transparent at these wavelengths \citep{Dehghanian19}. But the coronal lines can very well differentiate between the obscured and unobscured SED through constraints on the gas density. The value of $\log(n_\mathrm{H}/\mathrm{cm}^{-3}) \sim 9$ in the unobscured case (model 8) is what is expected of the material in the dusty torus.

Whilst the shapes of the \silvi\ wing and hot dust light-curves are strikingly similar (Fig.~\ref{fig:dust_wings}), it is noteworthy that the amplitude of the \silvi\ wing variability is substantially greater than that of the dust.
The wing varies by more than an order of magnitude whereas the dust flux variations are only at the $\pm25$~per~cent level.
Stronger variability of the coronal lines in \ngc\ compared with the dust on timescales of several years was also reported by \cite{Landt15b}. This behaviour is expected if the coronal line wing emission varies mainly in dependence on the compact X-ray source which in turn varies with a higher amplitude and on shorter timescales than the UV/optical emission that heats the dust \citep{Edelson2015}.

Both our spectroscopic analysis and photoionisation models point to the inner face of the torus being the origin of the coronal line wings.
But we stress that we cannot claim that our \cloudy\ models present unique and definitive solutions and there are several caveats to bear in mind, such as, e.g.\ the assumption of particular SEDs non-contemporaneous with our emission line observations and the inexhaustive exploration of the parameter space. Nonetheless, these models serve to illustrate that our proposed solution for the origin of the coronal line wings is physically plausible.

\subsection{Links to outflowing absorption systems} 
We interpreted the blueshifts of the coronal line cores as evidence of emission from outflowing gas.
Outflows in \ngc\ have been detected via absorption lines in other wavebands including X-rays (\citealt{Ebrero16}), UV (\citealt{Arav15}) and near-IR (\citealt{Wildy21}).
Some of the X-ray warm absorber components in \ngc\ are related to the UV absorbers (e.g.\ \citealt{Ebrero16}; \citealt{Arav15}) and \cite{Wildy21} associated several narrow He\,\textsc{i}~$\lambda1.08$~\um absorption features with corresponding UV absorbers.
Similarities between the inferred properties of the coronal line region and warm absorber (both are thought to originate in a highly-ionised, outflowing gas on scales intermediate between the BLR and NLR) have prompted further investigation into the possibility that they are in fact the same medium.
\cite{Por99} used photoionisation models to demonstrate that coronal emission lines could indeed form within the warm absorber.
They found that very high gas densities ($\log[n_\mathrm{H}/\mathrm{cm}^{-3}]\gtrsim10$) were required  to avoid overproduction of the coronal lines relative to observations and that the gas was located on scales similar to the BLR.
It is therefore pertinent to investigate whether the near-IR emission features seen in the outflow of \ngc\ (i.e. the coronal lines) can be associated with the gas responsible for the observed absorption features.

The associations of X-ray, UV and near-IR absorption features was based on the gaseous kinematics (i.e.\ similar velocity shifts), with further consideration made of the gas properties (density, ionisation parameter).
The velocity shifts we determined for the coronal lines ($\approx-125$~\kms\ for \se, \sn\ and \sit; $\approx-380$~\kms\ for \silvi) do not closely match those reported for any of the absorption features mentioned above.
Additionally, the majority of these absorption features arise at much greater distances from the nucleus than we infer for the coronal lines.
The estimated location of the UV absorption clouds is at the NLR radius and above ($\gtrsim2$~pc or $\log[R/\mathrm{cm}]\gtrsim18.8$), whereas the coronal lines are likely to be produced interior to the NLR.
The highest-ionisation components of the X-ray warm absorber are located on smaller scales ($\log[R/\mathrm{cm}]\gtrsim18$), similar to what we infer for the coronal lines, although these components have much higher outflow velocities (blueshifts of $\approx250$--$1200$~\kms) and much narrower widths ($\sigma_v\approx20$--200~\kms) than we observe in the coronal lines.
Our photoionisation modelling with \cloudy\ suggests that the coronal lines observed in \ngc\ form in gas of densities $\log(n_\mathrm{H}/\mathrm{cm}^{-3})\sim7$--9, much lower density than the $\log(n_\mathrm{H}/\mathrm{cm}^{-3})\gtrsim10$ required by \cite{Por99} but also much higher than the range of densities calculated for the warm absorber in \ngc\ ($\log[n_\mathrm{H}/\mathrm{cm}^{-3}]\sim3$--5: \citealt{Ebrero16}).

In summary, although the outflow velocities of the coronal line gas are of the same order of magnitude as those of some of the observed absorbers ($\sim$few hundred \kms), we see little evidence of an association between the coronal line gas and the specific X-ray/UV absorption components previously reported in \ngc. Then, if the coronal line gas and the absorbers are indeed parts of the same wind, they must arise from widely different clumps and so are unlikely to be exactly cospatial.


\section{Summary and conclusions} \label{sec:conclusions}
We have performed the first intensive study of the variability of the near-infrared coronal lines in an AGN.
The near-infrared spectroscopic monitoring campaign of \cite{Landt19} reveals a complex and multi-component coronal line region.
From measurements of the high-S/N mean spectrum of the campaign, we observe some general trends which have been observed in previous studies of AGN  coronal lines.
The narrow cores of the coronal lines are broader than, and blueshifted with respect to, the lower-ionisation forbidden lines\textcolor{black}{.  The lack of strong variability of the coronal line cores in the year-long campaign suggests the majority of the line-emitting gas exists on scales $\gtrsim1$~light year, whilst the line widths suggest the gas is more compact than the standard NLR ($\lesssim3$~light years).} 
These findings suggest that the narrow near-IR coronal lines in \ngc\ form in an outflow just interior to the inner NLR, although our photoionisation models show that coronal line emission originating closer to the BLR is possible.
An approximate correlation between the FWHM and ionisation potential of the coronal line cores implies a stratification of line emitting gas, with the lowest-ionisation coronal line \silvi\ line forming at a greater distance from the nucleus than the other lines.
The greater blueshift and narrower line profile of \silvi\ with respect to the higher-ionisation lines \se\ and \sit\ can be understood if this stratified wind is accelerating.
We therfore propose that the narrow coronal line cores form in a stratified, accelerating outflow from the nucleus of \ngc\ and this emission occurs predominantly just interior to the standard NLR.

In addition to narrow cores, the two lowest-ionisation coronal lines \se\ and \silvi\ show extended wings, particularly on the red side of the core.
The line widths, strong flux variability on a timescale of $\sim$months and photoionisation predictions all indicate that the wings are produced at a radius $\log(R/\mathrm{cm})\approx17.3$ from the nucleus. 
This radius is exactly coincident with the inner dust radius determined by \cite{Landt19}.
We therefore associate these wings with an X-ray heated wind of material evaporating from the inner face of the torus.

This study has revealed the complex and multi-faceted nature of the coronal line region in an AGN.
The recent launch of the \textit{James Webb Space Telescope} (\textit{JWST}) mean it is an opportune time to investigate this complexity.
The high sensitivity and spectral resolution of \textit{JWST}'s instruments will enable us to study infrared coronal lines of AGN in much greater detail.

\section*{Acknowledgements}
DK acknowledges support from the Czech Science Foundation project No.\ 19-05599Y, funding from the Czech Academy of Sciences via the PPLZ program and the receipt of a UK Science and Technology Facilities Council (STFC) studentship (ST/N50404X/1). 
HL acknowledges a Daphne Jackson Fellowship sponsored by the STFC. 
MJW acknowledges support from STFC grants ST/P000541/1 and ST/T000244/1.
G.J.F. and M.D. acknowledge support from NSF (1816537, 1910687), NASA (ATP 17-ATP17-0141, 19-ATP19-0188), and STScI (HST-AR- 15018 and HST-GO16196.003-A).

\section*{Data availability}
The raw data underlying this work are publicly available at the NASA IRTF Archive hosted by the NASA/IPAC Infrared Science Archive (\url{https://irsa.ipac.caltech.edu}).
The processed data are available on request from the second author. 




\bibliographystyle{mnras}
\bibliography{refs} 



\appendix
\section{Taking a look into a typical coronal line emitting cloud}
\begin{figure}
    \centering
    \includegraphics[width=\columnwidth]{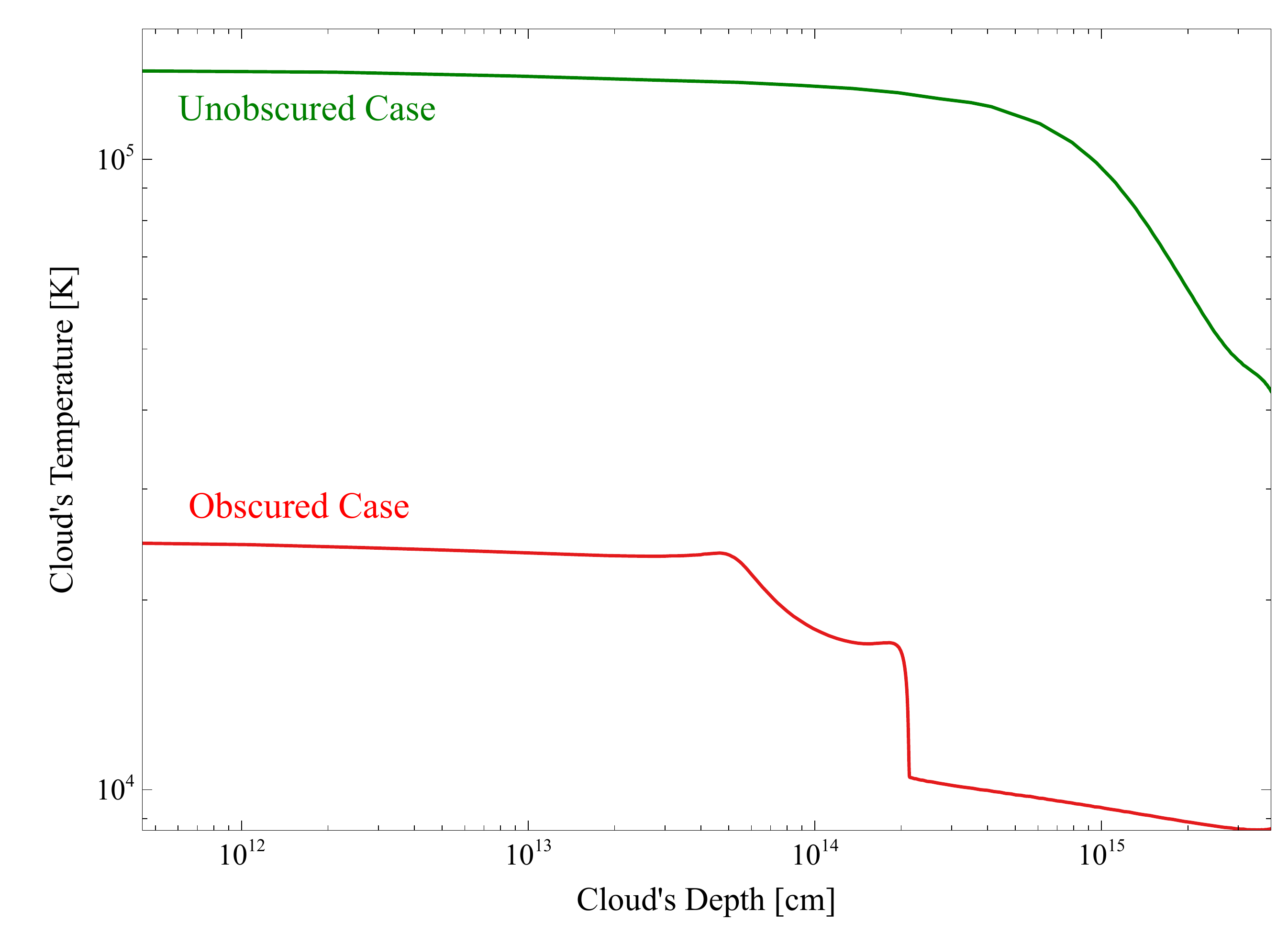}
    \caption{The variation of the temperature versus the depth into the coronal cloud is shown.  }
    \label{fig:Te}
\end{figure}

Fig.~\ref{fig:Te} shows the temperature inside a coronal line emitting cloud for both
unobscured (green) and obscured (red) cases.  
The model belongs to a cloud with
a column density of $N_\mathrm{H}=10^{23}$~cm$^{-2}$, hydrogen density of $n_\mathrm{H}=10^{7.4}$~cm$^{-3}$, located at $R=10^{17.3}$~cm from the
source. 
Much of our previous work rejected clouds with temperatures
greater than the $\sim10^5$~K peak in the cooling function since they produce
negligible optical or IR emission.  
However, the large column density
clouds have enough extinction to produce cooler gas near the shielded
side of the cloud, so we include these for completeness.  
These clouds occupy the lower left quadrant of the density-radius plane shown in Fig.~\ref{fig:cloudy}.

\bsp	
\label{lastpage}
\end{document}